# MIMO B-MAC Interference Network Optimization under Rate Constraints by Polite Water-filling and Duality

An Liu, Youjian (Eugene) Liu, Haige Xiang, Wu Luo

## Abstract

We take two new approaches to design efficient algorithms for transmitter optimization under rate constraints, to guarantee the Quality of Service in general MIMO interference networks, which is a combination of multiple interfering broadcast channels (BC) and multiaccess channels (MAC) and is named B-MAC Networks. Two related optimization problems, maximizing the minimum of weighted rates under a sum-power constraint and minimizing the sum-power under rate constraints, are considered. The first approach takes advantage of existing efficient algorithms for SINR problems by building a bridge between rate and SINR through the design of optimal mappings between them. The approach can be applied to other optimization problems as well. The second approach employs polite water-filling, which is the optimal network version of water-filling that we recently found. It replaces most generic optimization algorithms currently used for networks and reduces the complexity while demonstrating superior performance even in non-convex cases. Both centralized and distributed algorithms are designed and the performance is analyzed in addition to numeric examples.

The work was supported in part by NSFC Grant No.60972008, and in part by US-NSF Grant CCF-0728955 and ECCS-0725915. An Liu (Email: wendaol@pku.edu.cn), Haige Xiang, and Wu Luo are with the State Key Laboratory of Advanced Optical Communication Systems & Networks, School of EECS, Peking University, China. Youjian Liu is with the Department of Electrical, Computer, and Energy Engineering, University of Colorado at Boulder, USA. The corresponding author is Wu Luo (Email: luow@pku.edu.cn).

# I. INTRODUCTION

## A. System Setup and Problem Statement

We study the optimization under rate constraints for general multiple-input multiple-output (MIMO) interference networks, named MIMO B-MAC networks [1], where each transmitter may send data to multiple receivers and each receiver may collect data from multiple transmitters. Consequently, the network is a combination of multiple interfering broadcast channels (BC) and multiaccess channels (MAC). It includes BC, MAC, interference channels, X channels [2], [3], X networks [4] and most practical wireless networks, such as cellular networks, WiFi networks, DSL, as special cases. We assume Gaussian input and that each interference is either completely cancelled or treated as noise. A wide range of interference cancellation is allowed, from no cancellation to any cancellation specified by a valid binary *coupling matrix* of the data links. For example, simple linear receivers, dirty paper coding (DPC) [5] at transmitters, and/or successive interference cancellation (SIC) at receivers may be employed.

Two optimization problems are considered for guaranteeing the Quality of Service (QoS), where each data link has a target rate. The feasibility of the target rates is determined by a feasibility optimization problem (**FOP**) which maximizes the minimum of scaled rates of all links, where the scale factors are the inverse of the target rates. **FOP** can be used in admission control. If the target rates are feasible, the system tries to operate at minimum total transmission power in order to prolong total battery life and to reduce the total interference to other networks by solving the sum power minimization problem (**SPMP**) under the rate constraints.

## B. Related Works

The SINR version of **FOP** and **SPMP** under SINR constraints have been well studied, e.g., [6]–[10] using SINR duality [11]–[14], which means that if a set of SINRs is achievable in the forward links, then the same SINRs can be achieved in the reverse links when the set of transmit and receive beamforming vectors are fixed. Thus, optimizing the transmit vectors of the forward links is equivalent to the simpler problem of optimizing the receive vectors in the reverse links. However, these algorithms lack the following. 1) They cannot be directly used to solve **FOP** and **SPMP** under rate constraints because the optimal number of beams for each link and the power/rate allocation over these beams are unknown; 2) Except for [9], interference cancellation is not considered; 3) The optimal encoding and decoding order when interference cancellation is employed is not solved.

Considering interference cancellation and encoding/decoding order, the **FOP** and **SPMP** for MIMO BC/MAC have been completely solved in [15] by converting them to convex weighted sum-rate max-imization problems for MAC. The complexity is very high because the steepest ascent algorithm for the weighted sum-rate maximization needs to be solved repeatedly for each weight vector searched by the ellipsoid algorithm. A high complexity algorithm that can find the optimal encoding/decoding order for MISO BC/SIMO MAC is proposed in [16] that needs several inner and outer iterations. A heuristic low-complexity algorithm in [16] finds the near-optimal encoding/decoding order for **SPMP** by observing that the optimal solution of **SPMP** must be the optimal solution of some weighted sum-rate maximization problem, in which the weight vector can be found and used to determine the decoding order.

*C. Contribution*

The contribution of the paper is as follows.

- *Rate-SINR Conversion*: One difficulty of solving the problems is the joint optimization of beamforming matrices of all links. One approach is to decompose a link to multiple single-input single-output (SISO) streams and optimize the beamforming vectors by SINR duality, if a bridge between rate and SINR can be built to determine the optimal number of streams and rate/power allocation among the streams. In Section IV-A, we show that any Pareto rate point of an achievable rate region can be mapped to Pareto SINR points of the achievable SINR region through two optimal and simple mappings that produce equal rate and equal power streams respectively. The significance of this result is that it offers a method to convert the rate problems to SINR problems.

- *SINR based Algorithms*: Using the above result, we take advantage of existing algorithms for SINR problems to solve **FOP** and **SPMP** under rate constraints in Section IV-B and provide optimality analysis in Section IV-C.

- *Polite Water-filling based Algorithms*: Another approach is to directly solve for the beamforming matrices. For the convex problem of MIMO MAC, steepest ascent algorithm is used except for the special case of sum-rate optimal points, where iterative water-filling can be employed [17]–[19]. The B-MAC network problems are non-convex in general and thus, better algorithms, like water-filling, than the steepest ascent algorithm is highly desirable. However, directly applying traditional water-filling is far from optimal [20]–[22]. In [1], we recently found the long sought optimal network version of water-filling, polite water-filling, which is the optimal input structure of any Pareto rate point, not only the sum-rate optimal point, of the achievable region of a MIMO B-MAC network. In Section IV-D, polite water-filling based algorithms are designed that have low complexity and high performance.

- *Distributed Algorithm*: In a network, it is highly desirable to use distributed algorithms. The polite

water-filling based algorithm is well suited for distributed implementation, which is shown in Section IV-E, where each node only needs to estimate/exchange the local channel state information (CSI) but the performance is almost the same as that of the centralized algorithm.

- *Optimization of Encoding and Decoding Orders*: Another difficulty is to find the optimal encoding/decoding order when interference cancellation like DPC/SIC are employed. Again, polite water-filling proves useful in Section IV-F because the water-filling levels of the links can be used to identify the optimal encoding/decoding order for BC/MAC and pseudo-BC/MAC defined later.

The rest of the paper is organized as follows. Section II defines the achievable rate region and formulates the problems. Section III summarizes the preliminaries on duality and polite water-filling. Section IV presents the efficient centralized and distributed algorithms. The performance of the algorithms is demonstrated by simulation in Section V. The conclusion is given in Section VI. Due to the limited space, some proofs and algorithm description are shortened. The details can be found in the technical report [23].

## II. SYSTEM MODEL AND PROBLEM FORMULATION

### A. Definition of the Achievable Rate Region

We consider a MIMO B-MAC interference network consisting of $L$ data links. Let $T_l$ and $R_l$ denote the virtual transmitter and receiver of link $l$ equipped with $L_{T_l}$ transmit antennas and $L_{R_l}$ receive antennas respectively. The received signal at $R_l$ is $\mathbf{y}_l = \sum_{k=1}^{L} \mathbf{H}_{l,k} \mathbf{x}_k + \mathbf{w}_l$, where $\mathbf{x}_k \in \mathbb{C}^{L_{T_k} \times 1}$ is the transmit signal of link $k$ and is assumed to be circularly symmetric complex Gaussian; $\mathbf{H}_{l,k} \in \mathbb{C}^{L_{R_l} \times L_{T_k}}$ is the channel matrix between $T_k$ and $R_l$; and $\mathbf{w}_l \in \mathbb{C}^{L_{R_l} \times 1}$ is a circularly symmetric complex Gaussian noise vector with zero mean and identity covariance matrix.

To handle a wide range of interference cancellation possibilities, we define a coupling matrix $\mathbf{\Phi} \in \mathbb{R}_+^{L \times L}$ as a function of the interference cancellation scheme [1]. It specifies whether interference is completely cancelled or treated as noise: if $\mathbf{x}_k$, after interference cancellation, still causes interference to $\mathbf{x}_l$, $\mathbf{\Phi}_{l,k} = 1$ and otherwise, $\mathbf{\Phi}_{l,k} = 0$. The coupling matrices valid for the results of this paper are those for which there exists a transmission and receiving scheme such that each signal is decoded and completely cancelled by no more than one receiver [1]. In Fig. 1, we give an example of a B-MAC network employing DPC and SIC. When no data is transmitted over link 4 and 5, the following $\mathbf{\Phi}^a, \mathbf{\Phi}^b$ are valid coupling matrices for link $1, 2, 3$ under the corresponding encoding and decoding orders: *a*. $\mathbf{x}_1$ is encoded after $\mathbf{x}_2$ and $\mathbf{x}_2$ is decoded after $\mathbf{x}_3$; b. $\mathbf{x}_2$ is encoded after $\mathbf{x}_1$ and $\mathbf{x}_2$ is decoded after $\mathbf{x}_3$.

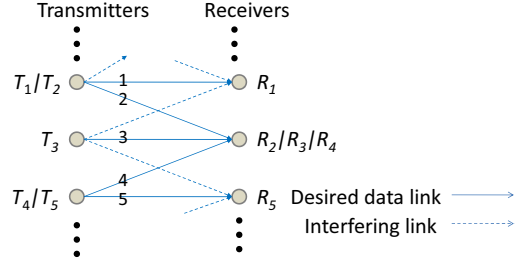

Figure 1. Example of a B-MAC network

$$\boldsymbol{\Phi}^a = \begin{bmatrix} 0 & 0 & 1 \\ 1 & 0 & 0 \\ 1 & 1 & 0 \end{bmatrix}, \boldsymbol{\Phi}^b = \begin{bmatrix} 0 & 1 & 1 \\ 0 & 0 & 0 \\ 1 & 1 & 0 \end{bmatrix}.$$

Note that when DPC and SIC are combined, an interference may not be fully cancelled under a specific encoding and decoding order. Such case cannot be described by the coupling matrix of 0's and 1's defined above. But a valid coupling matrix can serve for an upper or lower bound. See more discussion in [1].

The achievable regions in this paper refer to the following. Note that $\boldsymbol{\Phi}_{l,l} = 0$ by definition. The interference-plus-noise covariance matrix of the $l^{\text{th}}$ link is

$$\boldsymbol{\Omega}_l = \mathbf{I} + \sum_{k=1}^{L} \boldsymbol{\Phi}_{l,k} \mathbf{H}_{l,k} \boldsymbol{\Sigma}_k \mathbf{H}_{l,k}^\dagger, \quad (1)$$

where $\boldsymbol{\Sigma}_k$ is the covariance matrix of $\mathbf{x}_k$. We denote all the covariance matrices as $\boldsymbol{\Sigma}_{1:L} = (\boldsymbol{\Sigma}_1, \boldsymbol{\Sigma}_2, ..., \boldsymbol{\Sigma}_L)$. Then the achievable mutual information (rate) of link $l$ is given by a function of $\boldsymbol{\Sigma}_{1:L}$ and $\boldsymbol{\Phi}$

$$\mathcal{I}_l (\boldsymbol{\Sigma}_{1:L}, \boldsymbol{\Phi}) = \log \left| \mathbf{I} + \mathbf{H}_{l,l} \boldsymbol{\Sigma}_l \mathbf{H}_{l,l}^\dagger \boldsymbol{\Omega}_l^{-1} \right|. \quad (2)$$

*Definition 1:* The *Achievable Rate Region* with a fixed coupling matrix $\boldsymbol{\Phi}$ and sum power constraint $P_T$ is defined as

$$\mathcal{R}_{\boldsymbol{\Phi}} (P_T) \triangleq \bigcup_{\boldsymbol{\Sigma}_{1:L}: \sum_{l=1}^{L} \text{Tr}(\boldsymbol{\Sigma}_l) \leq P_T} \left\{ \mathbf{r} \in \mathbb{R}_+^L : \quad (3) \right.$$
$$\left. r_l \leq \mathcal{I}_l (\boldsymbol{\Sigma}_{1:L}, \boldsymbol{\Phi}), 1 \leq l \leq L \right\}.$$

The algorithms rely on the duality between the forward and reverse links of a B-MAC network. The reverse links are obtained by reversing the transmission direction and replacing the channel matrices by their conjugate transposes. The coupling matrix for the reverse links is the transpose of that for the forward links. We use the notation ˆ to denote the corresponding terms in the reverse links. The interference-plus-noise covariance matrix of reverse link $l$ is

$$\hat{\mathbf{\Omega}}_l = \mathbf{I} + \sum_{k=1}^{L} \mathbf{\Phi}_{k,l} \mathbf{H}_{k,l}^\dagger \hat{\mathbf{\Sigma}}_k \mathbf{H}_{k,l}, \quad (4)$$

and the rate of reverse link $l$ is given by $\hat{\mathcal{I}}_l \left( \hat{\mathbf{\Sigma}}_{1:L}, \mathbf{\Phi}^T \right) = \log \left| \mathbf{I} + \mathbf{H}_{l,l}^\dagger \hat{\mathbf{\Sigma}}_l \mathbf{H}_{l,l} \hat{\mathbf{\Omega}}_l^{-1} \right|$.

## B. Problem Formulation

This paper concerns the feasibility optimization problem (**FOP**) and the sum power minimization problem (**SPMP**) under Quality of Service (QoS) constraints in terms of target rates $\left[ \mathcal{I}_l^0 \right]_{l=1,...,L}$ for a B-MAC network with a given valid coupling matrix $\mathbf{\Phi}$:

$$\textbf{FOP}: \quad \max_{\mathbf{\Sigma}_{1:L}} \left( \min_{1 \leq l \leq L} \frac{\mathcal{I}_l \left( \mathbf{\Sigma}_{1:L}, \mathbf{\Phi} \right)}{\mathcal{I}_l^0} \right) \quad (5)$$

$$\text{s.t.} \quad \mathbf{\Sigma}_l \succeq 0, l = 1, ..., L \text{ and } \sum_{l=1}^{L} \text{Tr} \left( \mathbf{\Sigma}_l \right) \leq P_T,$$

where $P_T$ is the total power constraint;

$$\textbf{SPMP}: \quad \min_{\mathbf{\Sigma}_{1:L}} \sum_{l=1}^{L} \text{Tr} \left( \mathbf{\Sigma}_l \right) \quad (6)$$

$$\text{s.t.} \quad \mathcal{I}_l \left( \mathbf{\Sigma}_{1:L}, \mathbf{\Phi} \right) \geq \mathcal{I}_l^0, \mathbf{\Sigma}_l \succeq 0, l = 1, ..., L.$$

Solving **FOP** is the same as finding the intersection of vector $\left[ \mathcal{I}_l^0 \right]_{l=1,...,L}$ and the achievable region boundary, i.e., the optimal rate vector satisfies $\left[ \mathcal{I}_l \right]_{l=1,...,L} = \alpha \left[ \mathcal{I}_l^0 \right]_{l=1,...,L}$. If the target rates are feasible, the **SPMP** finds the minimum total power needed.

For the special case of DPC and SIC, the optimal coupling matrix $\mathbf{\Phi}$, or equivalently, the optimal encoding and/or decoding order of **FOP** and **SPMP** is partially solved in Section IV-F. We first focus on centralized algorithms. Then we give a distributed implementation of the algorithm for **SPMP** under additional individual maximum power constraints.

Although we focus on the sum power and white noise, the results can be directly applied to a larger class of problems with a single linear constraint $\sum_{l=1}^{L} \text{Tr} \left( \mathbf{\Sigma}_l \hat{\mathbf{W}}_l \right) \leq P_T$ in **FOP** (or objective function $\sum_{l=1}^{L} \text{Tr} \left( \mathbf{\Sigma}_l \hat{\mathbf{W}}_l \right)$ in **SPMP**) and/or colored noise with covariance $\mathrm{E} \left[ \mathbf{w}_l \mathbf{w}_l^\dagger \right] = \mathbf{W}_l$. Only variable changes $\mathbf{\Sigma}_l' = \hat{\mathbf{W}}_l^{\frac{1}{2}} \mathbf{\Sigma}_l \hat{\mathbf{W}}_l^{\frac{1}{2}}$ and $\mathbf{W}_k^{-\frac{1}{2}} \mathbf{H}_{k,l} \hat{\mathbf{W}}_l^{-\frac{1}{2}}$ are needed, where $\hat{\mathbf{W}}_l$ and $\mathbf{W}_k$ are positive definite for meaningful cases[1]. The single linear constraint appears in Lagrange functions for problems with multiple linear constraints [24], [25], and thus, the results in this paper serve as the basis to solve them [26].

---

[1] For random channels, singular $\hat{\mathbf{W}}_l$ or $\mathbf{W}_l$ will result in infinite power and/or rate with probability one.

## III. PRELIMINARIES

The algorithms are based on SINR duality, e.g., [10], rate duality, and polite water-filling [1]. They are reviewed below.

### A. SINR Duality for MIMO B-MAC Networks

The achievable rate region defined in (3) can be achieved by the following spatial multiplexing scheme. We define the *Decomposition of a MIMO Link into Multiple SISO Data Streams* as, for link $l$ and $M_l \geq \text{Rank}(\Sigma_l)$, finding a precoding matrix $\dot{\mathbf{T}}_l = \left[\sqrt{p_{l,1}}\mathbf{t}_{l,1}, ..., \sqrt{p_{l,M_l}}\mathbf{t}_{l,M_l}\right]$ satisfying

$$\Sigma_l = \dot{\mathbf{T}}_l \dot{\mathbf{T}}_l^\dagger = \sum_{m=1}^{M_l} p_{l,m} \mathbf{t}_{l,m} \mathbf{t}_{l,m}^\dagger, \quad (7)$$

where $\mathbf{t}_{l,m} \in \mathbb{C}^{L_{T_l} \times 1}$ is a transmit vector with $\|\mathbf{t}_{l,m}\| = 1$; and $\mathbf{p} = [p_{1,1}, ..., p_{1,M_1}, ..., p_{L,1}, ..., p_{L,M_L}]^T$ are the transmit powers. Without loss of generality, we assume the intra-signal decoding order is that the $m^{\text{th}}$ stream is the $m^{\text{th}}$ to be decoded and cancelled. The receive vector for the $m^{th}$ stream of link $l$ is obtained by the MMSE filtering as

$$\mathbf{r}_{l,m} = \alpha_{l,m} \left( \sum_{i=m+1}^{M_l} \mathbf{H}_{l,l} p_{l,i} \mathbf{t}_{l,i} \mathbf{t}_{l,i}^\dagger \mathbf{H}_{l,l}^\dagger + \Omega_l \right)^{-1} \mathbf{H}_{l,l} \mathbf{t}_{l,m}, \quad (8)$$

where $\alpha_{l,m}$ is chosen such that $\|\mathbf{r}_{l,m}\| = 1$. This is referred to as MMSE-SIC receiver in this paper.

For each stream, one can calculate its SINR. Let the collections of transmit and receive vectors be

$$\mathbf{T} = [\mathbf{t}_{l,m}]_{m=1,...,M_l, l=1,...,L}, \quad (9)$$

$$\mathbf{R} = [\mathbf{r}_{l,m}]_{m=1,...,M_l, l=1,...,L}. \quad (10)$$

Define the $\left(\sum_{i=1}^{l-1} M_i + m\right)^{\text{th}}$ row and $\left(\sum_{i=1}^{k-1} M_i + n\right)^{\text{th}}$ column of the cross-talk matrix $\boldsymbol{\Psi}(\mathbf{T}, \mathbf{R}) \in \mathbb{R}_+^{\sum_l M_l \times \sum_l M_l}$ [8] as

$$\boldsymbol{\Psi}_{l,m}^{k,n} = \begin{cases} 0 & k = l \text{ and } m \geq n, \\ \left|\mathbf{r}_{l,m}^\dagger \mathbf{H}_{l,l} \mathbf{t}_{l,n}\right|^2 & k = l, \text{ and } m < n, \\ \Phi_{l,k} \left|\mathbf{r}_{l,m}^\dagger \mathbf{H}_{l,k} \mathbf{t}_{k,n}\right|^2 & \text{otherwise.} \end{cases} \quad (11)$$

Then the SINR for the $m^{th}$ stream of link $l$ is

$$\gamma_{l,m}(\mathbf{T}, \mathbf{R}, \mathbf{p}) = \frac{p_{l,m} \left|\mathbf{r}_{l,m}^\dagger \mathbf{H}_{l,l} \mathbf{t}_{l,m}\right|^2}{1 + \sum_{k=1}^{L} \sum_{n=1}^{M_k} p_{k,n} \boldsymbol{\Psi}_{l,m}^{k,n}}. \quad (12)$$

Such decomposition of data to streams with MMSE-SIC receiver is information lossless [27], i.e., the sum-rate of all streams of link $l$ is equal to the mutual information in (2).

In the reverse links, we can obtain SINRs using $\mathbf{R}$ as transmit vectors and $\mathbf{T}$ as receive vectors. The transmit powers are denoted as $\mathbf{q} = [q_{1,1}, ..., q_{1,M_1}, ..., q_{L,1}, ..., q_{L,M_L}]^T$. The intra-signal decoding order is the opposite to that of the forward link, i.e., the $m^{\text{th}}$ stream is the $m^{\text{th}}$ last to be decoded and cancelled. Then the SINR for the $m^{\text{th}}$ stream of reverse link $l$ is

$$\hat{\gamma}_{l,m}(\mathbf{R}, \mathbf{T}, \mathbf{q}) = \frac{q_{l,m} \left| \mathbf{t}_{l,m}^{\dagger} \mathbf{H}_{l,l}^{\dagger} \mathbf{r}_{l,m} \right|^2}{1 + \sum_{k=1}^{L} \sum_{n=1}^{M_k} q_{k,n} \mathbf{\Psi}_{k,n}^{l,m}}. \tag{13}$$

For simplicity, we will use $\{\mathbf{T}, \mathbf{R}, \mathbf{p}\}$ ($\{\mathbf{R}, \mathbf{T}, \mathbf{q}\}$) to denote the transmission and reception strategy described above in the forward (reverse) links.

The achievable SINR regions of the forward and reverse links are the same. Define the achievable SINR regions $\mathcal{T}_{\mathbf{\Phi}}(P_T)$ and $\hat{\mathcal{T}}_{\mathbf{\Phi}^T}(P_T)$ as the set of all SINRs that can be achieved under the sum power constraint $P_T$ in the forward and reverse links respectively. For a given set of SINR values $\gamma^0 = \left[\gamma_{l,m}^0\right]_{m=1,...,M_l, l=1,...,L}$, define a diagonal matrix $\mathbf{D}(\mathbf{T}, \mathbf{R}, \gamma^0) \in \mathbb{R}_+^{\sum_l M_l \times \sum_l M_l}$ where the $\left(\sum_{i=1}^{l-1} M_i + m\right)^{\text{th}}$ diagonal element is

$$\mathbf{D}_{\sum_{i=1}^{l-1} M_i + m, \sum_{i=1}^{l-1} M_i + m} = \gamma_{l,m}^0 / \left| \mathbf{r}_{l,m}^{\dagger} \mathbf{H}_{l,l} \mathbf{t}_{l,m} \right|^2. \tag{14}$$

We restate the SINR duality, e.g. [10], as follows.

*Lemma 1:* If a set of SINRs $\gamma^0$ is achieved by the transmission and reception strategy $\{\mathbf{T}, \mathbf{R}, \mathbf{p}\}$ with $\|\mathbf{p}\|_1 = P_T$ in the forward links, then $\gamma^0$ is also achievable in the reverse links with $\{\mathbf{R}, \mathbf{T}, \mathbf{q}\}$, where $\mathbf{q}$ satisfies $\|\mathbf{q}\|_1 = P_T$ and is given by

$$\mathbf{q} = \left(\mathbf{D}^{-1}(\mathbf{T}, \mathbf{R}, \gamma^0) - \mathbf{\Psi}^T(\mathbf{T}, \mathbf{R})\right)^{-1} \mathbf{1}. \tag{15}$$

And thus, one has $\mathcal{T}_{\mathbf{\Phi}}(P_T) = \hat{\mathcal{T}}_{\mathbf{\Phi}^T}(P_T)$.

## B. Rate Duality

We give the rate duality under a linear constraint $\sum_{l=1}^{L} \text{Tr}\left(\mathbf{\Sigma}_l \hat{\mathbf{W}}_l\right) \leq P_T$ and/or colored noise with covariance $\text{E}\left[\mathbf{w}_l \mathbf{w}_l^{\dagger}\right] = \mathbf{W}_l$ [1]. For convenience, let

$$\left([\mathbf{H}_{l,k}], \sum_{l=1}^{L} \text{Tr}\left(\mathbf{\Sigma}_l \hat{\mathbf{W}}_l\right) \leq P_T, [\mathbf{W}_l]\right), \tag{16}$$

denote the network with channel matrices $[\mathbf{H}_{l,k}]$. Then the reverse links (dual network) is given by

$$\left( \left[ \mathbf{H}_{k,l}^{\dagger} \right], \sum_{l=1}^{L} \text{Tr}\left( \hat{\mathbf{\Sigma}}_l \mathbf{W}_l \right) \leq P_T, \left[ \hat{\mathbf{W}}_l \right] \right). \tag{17}$$

*Theorem 1:* If $\mathbf{\Sigma}_{1:L}$ achieves certain rates and satisfies the linear constraint in network (16), its covariance transformation $\hat{\mathbf{\Sigma}}_{1:L}$ calculated by

$$\hat{\mathbf{\Sigma}}_l = \sum_{m=1}^{M_l} q_{l,m} \mathbf{r}_{l,m} \mathbf{r}_{l,m}^{\dagger}, l = 1, ..., L, \tag{18}$$

and (8,15) achieves equal or larger rates in the reverse links (17) under the linear constraint $\sum_{l=1}^{L} \text{Tr}\left(\hat{\mathbf{\Sigma}}_l \mathbf{W}_l\right) = \sum_{l=1}^{L} \text{Tr}\left(\mathbf{\Sigma}_l \hat{\mathbf{W}}_l\right) \leq P_T$. Therefore, the achievable rate regions of the forward and reverse links of a B-MAC network are the same.

### C. Polite Water-filling

In [1], we showed that the Pareto optimal input covariance matrices have a *polite water-filling structure*. It generalizes the well known optimal single user water-filling structure to networks.

*Definition 2:* Given input covariance matrices $\mathbf{\Sigma}_{1:L}$, obtain its covariance transformation $\hat{\mathbf{\Sigma}}_{1:L}$ by (18) and calculate the interference-plus-noise covariance matrices $\mathbf{\Omega}_{1:L}$ and $\hat{\mathbf{\Omega}}_{1:L}$. Pre- and post- whiten the channel $\mathbf{H}_{l,l}$ to produce an equivalent single user channel $\bar{\mathbf{H}}_l = \mathbf{\Omega}_l^{-1/2} \mathbf{H}_{l,l} \hat{\mathbf{\Omega}}_l^{-1/2}$. Define $\mathbf{Q}_l \triangleq \hat{\mathbf{\Omega}}_l^{1/2} \mathbf{\Sigma}_l \hat{\mathbf{\Omega}}_l^{1/2}$ as the equivalent input covariance matrix of link $l$. Matrix $\mathbf{\Sigma}_l$ is said to possess a *polite water-filling structure* if $\mathbf{Q}_l$ is a water-filling over $\bar{\mathbf{H}}_l$, i.e.,

$$\begin{aligned} \mathbf{Q}_l &= \mathbf{G}_l \mathbf{D}_l \mathbf{G}_l^{\dagger}, \\ \mathbf{D}_l &= \left( \nu_l \mathbf{I} - \mathbf{\Delta}_l^{-2} \right)^+. \end{aligned} \tag{19}$$

where $\nu_l \geq 0$ is the *polite water-filling level*; the equivalent channel $\bar{\mathbf{H}}_l$'s thin singular value decomposition (SVD) is $\bar{\mathbf{H}}_l = \mathbf{F}_l \mathbf{\Delta}_l \mathbf{G}_l^{\dagger}$ with $\mathbf{F}_l \in \mathbb{C}^{L_{R_l} \times N_l}$, $\mathbf{G}_l \in \mathbb{C}^{L_{T_l} \times N_l}$, $\mathbf{\Delta}_l \in \mathbb{R}_{++}^{N_l \times N_l}$, and $N_l = \text{Rank}(\mathbf{H}_{l,l})$. If all $\mathbf{\Sigma}_l$'s possess the polite water-filling structure, then $\mathbf{\Sigma}_{1:L}$ is said to possess the polite water-filling structure.

*Theorem 2:* The input covariance matrices $\mathbf{\Sigma}_{1:L}$ of any Pareto rate point of the achievable rate region and its covariance transformation $\hat{\mathbf{\Sigma}}_{1:L}$ possess the polite water-filling structure.

We have the following even at a non-Pareto rate point [1].

*Theorem 3:* If one input covariance matrix $\mathbf{\Sigma}_l$ has the polite water-filling structure while other $\mathbf{\Sigma}_k, \hat{\mathbf{\Sigma}}_k$, $k \neq l$, are fixed, so does its covariance transformation $\hat{\mathbf{\Sigma}}_l$, i.e., $\hat{\mathbf{Q}}_l \triangleq \mathbf{\Omega}_l^{1/2} \hat{\mathbf{\Sigma}}_l \mathbf{\Omega}_l^{1/2}$ is a water-filling over the reverse equivalent channel $\bar{\mathbf{H}}_l^{\dagger} \triangleq \hat{\mathbf{\Omega}}_l^{-1/2} \mathbf{H}_{l,l}^{\dagger} \mathbf{\Omega}_l^{-1/2}$.

# IV. Optimization Algorithms

In this section, we present several related algorithms for the feasibility optimization problem (**FOP**) and the sum power minimization problem (**SPMP**). Algorithms for SINR version of **FOP** and **SPMP** have been designed in [8], [10]. To take advantage of them, we show how to map a Pareto point of the achievable rate region to a Pareto point of the SINR region in Section IV-A and then use SINR based Algorithm A and B to solve **FOP** and **SPMP** respectively in Section IV-B. The optimality of Algorithms A and B is studied in Section IV-C by examining the structure of the optimal solutions of **FOP** and **SPMP**. The optimal structure suggests that the rate constrained problems can be directly solved using Algorithm PR and PR1 in Section IV-D by polite water-filling. In a network, it is desirable to have distributed algorithms, for which Algorithm PRD is designed in Section IV-E. Finally, we design Algorithm O to improve the encoding and decoding orders for all of the above algorithms when DPC and SIC are employed.

## A. Rate-SINR Conversion

In order to find Pareto rate points of the achievable rate region by taking advantage of algorithms that find Pareto points of the SINR region, one needs to find a mapping from a Pareto rate point to a Pareto SINR point. But multiple SINR points can correspond to the same rate. The following two theorems give an equal SINR mapping and an equal power mapping without loss of optimality by choosing two decompositions of a MIMO link to multiple SISO data streams.

*Theorem 4:* For any input covariance matrices $\boldsymbol{\Sigma}_{1:L}$ achieving a rate point $[\mathcal{I}_l]_{l=1,...,L}$, there exists a decomposition $\boldsymbol{\Sigma}_l = \sum_{m=1}^{M_l} p_{l,m} \mathbf{t}_{l,m} \mathbf{t}_{l,m}^\dagger, l = 1, ..., L$, with any integer $M_l \geq \text{Rank}(\boldsymbol{\Sigma}_l)$, such that the corresponding transmission and MMSE-SIC reception strategy $\{\mathbf{T}, \mathbf{R}, \mathbf{p}\}$ achieves equal SINR for all streams of the same link, i.e., $\gamma_{l,m} = e^{\mathcal{I}_l/M_l} - 1, m = 1, ..., M_l, l = 1, ..., L$. Therefore uniform rate allocation over the streams of the same link will not lose optimality.

The proof and the algorithm to find the decomposition are given in appendix A.

*Corollary 1:* Let $M_l = \text{Rank}(\mathbf{H}_{l,l})$. An SINR point $\left[\gamma_{l,m} = e^{\mathcal{I}_l/M_l} - 1\right]_{m=1,...,M_l, l=1,...,L}$ is a Pareto boundary point in $\mathcal{T}_{\boldsymbol{\Phi}}(P_T)$, if and only if the rate point $[\mathcal{I}_l]_{l=1,...,L}$ is a Pareto rate point in $\mathcal{R}_{\boldsymbol{\Phi}}(P_T)$.

The following theorem proved in Appendix B shows that uniform power allocation across the streams within a link will also not lose optimality, which is useful in designing algorithms for individual power constraints and/or distributed optimization [28], [29].

*Theorem 5:* For any input covariance matrix $\boldsymbol{\Sigma}$, there exists a decomposition $\boldsymbol{\Sigma} = \sum_{m=1}^{M} p_m \mathbf{t}_m \mathbf{t}_m^\dagger$ such that the transmit power is uniformly allocated over the $M$ streams, i.e., $p_m = \text{Tr}(\boldsymbol{\Sigma})/M, \forall m$. Therefore uniform power allocation over the streams of the same link will not lose optimality.

## B. SINR based Algorithms

The results in Section IV-A serve as a bridge to solve the **FOP** or **SPMP** under rate constraints through the SINR optimization problems. First we show **FOP** is equivalent to the following SINR optimization problem in the sense of feasibility.

$$\textbf{EFOP}: \max_{\{\mathbf{T},\mathbf{R},\mathbf{p}\}} \min_{\substack{1 \leq m \leq M_l \\ 1 \leq l \leq L}} \frac{\gamma_{l,m}}{\gamma_l^0}, \text{ s.t. } \|\mathbf{p}\|_1 \leq P_T, \tag{20}$$

where $M_l = \text{Rank}(\mathbf{H}_{l,l})$ is the number of streams of link $l$; $\gamma_l^0 = e^{\mathcal{I}_l^0/M_l} - 1$ is the target SINR.

*Theorem 6:* The optimum of **FOP** (5) is not less than 1 if and only if the optimum of **EFOP** (20) is not less than 1.

*Proof:* If the optimum of **EFOP** is not less than 1, there exists a point $\left[\gamma_{l,m} \geq \gamma_l^0\right]_{m=1,...,M_l, l=1,...,L}$ in $\mathcal{T}_{\boldsymbol{\Phi}}(P_T)$. Then by Theorem 4, the rate point $\left[\mathcal{I}_l = M_l \log(1 + \gamma_{l,m}) \geq \mathcal{I}_l^0\right]_{l=1,...,L}$ lies in $\mathcal{R}_{\boldsymbol{\Phi}}(P_T)$, i.e., the optimum of **FOP** is not less than 1. The 'only if' part can be proved similarly. ∎

Similarly, it can be proved that **SPMP** is equivalent to the following SINR optimization problem.

$$\textbf{ESPMP}: \min_{\{\mathbf{T},\mathbf{R},\mathbf{p}\}} \|\mathbf{p}\|_1, \text{ s.t. } \gamma_{l,m} \geq \gamma_l^0, \begin{array}{l} 1 \leq m \leq M_l, \\ 1 \leq l \leq L. \end{array} \tag{21}$$

*Theorem 7:* If $\{\tilde{\mathbf{T}}, \tilde{\mathbf{R}}, \tilde{\mathbf{p}}\}$ is an optimum of **ESPMP** (21), the input covariance matrices $\tilde{\boldsymbol{\Sigma}}_{1:L}$ produced by $\tilde{\mathbf{T}}$ and $\tilde{\mathbf{p}}$ must be an optimum of **SPMP** (6). On the other hand, if $\tilde{\boldsymbol{\Sigma}}_{1:L}$ is an optimum of **SPMP**, there exists a decomposition leading to $\{\tilde{\mathbf{T}}, \tilde{\mathbf{R}}, \tilde{\mathbf{p}}\}$, which is an optimum of **ESPMP**.

Sketched below are Algorithm A that solves **FOP** by solving **EFOP** as in [8] and Algorithm B that solves **SPMP** by solving **ESPMP** as in [10]. The transmit and receive vectors $\mathbf{T}, \mathbf{R}$ and the forward and reverse power $\mathbf{p}, \mathbf{q}$ are iteratively optimized by switching between the forward and reverse links. For fixed $\mathbf{T}, \mathbf{p}$, the optimal receive vector is given by the MMSE-SIC receiver in (8). The SINR duality in Lemma 1 implies that the transmit vectors $\mathbf{T}$ can be optimized by switching to the reverse links and finding the optimal MMSE-SIC receive vectors for fixed $\mathbf{R}, \mathbf{q}$ as

$$\mathbf{t}_{l,m} = \beta_{l,m} \left( \sum_{i=1}^{m-1} q_{l,i} \mathbf{H}_{l,l}^\dagger \mathbf{r}_{l,i} \mathbf{r}_{l,i}^\dagger \mathbf{H}_{l,l} + \hat{\boldsymbol{\Omega}}_l \right)^{-1} \mathbf{H}_{l,l}^\dagger \mathbf{r}_{l,m} \tag{22}$$

where $\hat{\mathbf{\Omega}}_l$ is obtained from $\hat{\mathbf{\Sigma}}_k = \sum_{i=1}^{M_k} q_{k,i} \mathbf{r}_{k,i} \mathbf{r}_{k,i}^\dagger$ using (4), and the normalization factor $\beta_{l,m}$ is chosen such that $\|\mathbf{t}_{l,m}\| = 1$. The optimization for $\mathbf{p}$ and $\mathbf{q}$ is different for the two problems. For **EFOP**, with the optimal power $\tilde{\mathbf{p}}$, all the scaled SINRs in (20) should be equal to a constant $C_{\max}$ [8]. Therefore $\tilde{\mathbf{p}}$ satisfies the equations $\gamma_{l,m} = C_{\max} \gamma_l^0, \forall m, l$ and $\|\tilde{\mathbf{p}}\|_1 = P_T$, which together form an eigen-system [8]. Then $\tilde{\mathbf{p}}$ is the dominant eigenvector of the eigen-system with its last component scaled to one [8]. The optimal reverse link power $\tilde{\mathbf{q}}$ is obtained by solving a similar eigen-system. For **ESPMP**, in each iteration, after $\mathbf{R}$ ($\mathbf{T}$) is calculated from MMSE of the forward (reverse) link, $\mathbf{p}$ ($\mathbf{q}$) is adjusted according to power control

$$p_{l,m}^{(n+1)} = \frac{\gamma_l^0}{\gamma_{l,m}^{(n)}} p_{l,m}^{(n)}, \qquad (23)$$

$$q_{l,m}^{(n+1)} = \frac{\gamma_l^0}{\hat{\gamma}_{l,m}^{(n)}} q_{l,m}^{(n)}, \qquad (24)$$

where $p_{l,m}^{(n)}/q_{l,m}^{(n)}$ is the power after the $n^{\text{th}}$ update and $\gamma_{l,m}^{(n)}/\hat{\gamma}_{l,m}^{(n)}$ is the SINR after the $n^{\text{th}}$ update. Finally, the dual power $\mathbf{q}$ ($\mathbf{p}$) is calculated by the SINR duality in Lemma 1 to make the SINRs of the forward and reverse links equal with the same sum power. After obtaining $\mathbf{T}$ and $\mathbf{p}$, the corresponding input covariance matrices for **FOP** and **SPMP** can be easily obtained. The convergence of these algorithms are proved in [8], [10].

### C. Optimality Analysis for SINR based Algorithms

Algorithm A or B can find good solutions but may not find the optimum for general B-MAC networks. But we can still obtain insight of the problem and derive improved algorithms by finding the necessary conditions satisfied by the optimum.

To get rid of the non-differentiable objective function in **FOP** (5), we rewrite it into the following equivalent problem

$$\textbf{FOPa}: \quad \max_{\mathbf{\Sigma}_{1:L}} \quad \frac{\mathcal{I}_1(\mathbf{\Sigma}_{1:L}, \mathbf{\Phi})}{\mathcal{I}_1^0} \qquad (25)$$

$$\text{s.t.} \quad \frac{\mathcal{I}_l(\mathbf{\Sigma}_{1:L}, \mathbf{\Phi})}{\mathcal{I}_l^0} \geq \frac{\mathcal{I}_1(\mathbf{\Sigma}_{1:L}, \mathbf{\Phi})}{\mathcal{I}_1^0}, \forall l \neq 1$$

$$\mathbf{\Sigma}_l \succeq 0, l = 1, \cdots, L \text{ and } \sum_{l=1}^{L} \text{Tr}(\mathbf{\Sigma}_l) \leq P_T.$$

Then the following theorem holds.

*Theorem 8:* Necessity: If $\tilde{\mathbf{\Sigma}}_{1:L} = \left(\tilde{\mathbf{\Sigma}}_1, ..., \tilde{\mathbf{\Sigma}}_L\right)$ is an optimum of **FOPa** (25) or **SPMP** (6), it must satisfy the optimality conditions below:

1) It possesses the polite water-filling structure as in Definition 2.

2) The achieved rates must satisfy $\mathcal{I}_l \left( \tilde{\boldsymbol{\Sigma}}_{1:L}, \boldsymbol{\Phi} \right) = \alpha \mathcal{I}_l^0, l = 1, ..., L$, where for **FOPa**, $\alpha > 0$ is some constant; and for **SPMP**, $\alpha = 1$.

3) For **FOPa**, it satisfies $\sum_{l=1}^{L} \text{Tr} \left( \tilde{\boldsymbol{\Sigma}}_l \right) = P_T$.

Sufficiency for the KKT: If certain $\tilde{\boldsymbol{\Sigma}}_{1:L}$ satisfies the above optimality conditions for **FOPa** or **SPMP**, it satisfies the Karush–Kuhn–Tucker (KKT) conditions of **FOPa** or **SPMP**, and thus achieves a stationary point.

Sufficiency for the Optimum: If certain $\tilde{\boldsymbol{\Sigma}}_{1:L}$ satisfies the above optimality conditions for **FOPa** or **SPMP** and if the weighted sum rate $\sum_l^L \tilde{\nu}_l \mathcal{I}_l (\boldsymbol{\Sigma}_{1:L}, \boldsymbol{\Phi})$ is a concave function of $\boldsymbol{\Sigma}_{1:L}$, where $\tilde{\nu}_l$'s are the polite water-filling levels of $\tilde{\boldsymbol{\Sigma}}_{1:L}$, then $\tilde{\boldsymbol{\Sigma}}_{1:L}$ is the optimum of **FOPa** or **SPMP**.

We only give a sketch of the proof. It can be proved by contradiction that the optimums of **FOPa** and **SPMP** are Pareto optimal. By Theorem 2, they possess the polite water-filling structure. The second optimality condition can be proved by a proof similar to that of Lemma 1 in [10] for **ESPMP**. The third optimality condition follows from the fact that if the total transmit power is less that $P_T$, the extra power can be used to improve the rates of all links. The connection between the necessary optimality conditions and the KKT conditions can be proved by a proof similar to that of Theorem 13 in [1]. The sufficiency for the optimum for **FOPa** can be proved by the following two facts. 1) Suppose certain $\tilde{\boldsymbol{\Sigma}}_{1:L}$ satisfies the optimality conditions for **FOPa**. It can be shown that the optimum of **FOPa** is equal to the optimum of the following weighted sum rate maximization problem

$$\textbf{WSRMP:} \max_{\boldsymbol{\Sigma}_{1:L}} \sum_{l=1}^{L} \tilde{\nu}_l \mathcal{I}_l (\boldsymbol{\Sigma}_{1:L}, \boldsymbol{\Phi}) \qquad (26)$$

$$\text{s.t.} \boldsymbol{\Sigma}_l \succeq 0, l = 1, \cdots, L \text{ and } \sum_{l=1}^{L} \text{Tr}(\boldsymbol{\Sigma}_l) \leq P_T,$$

where $\{\tilde{\nu}_l\}$ are the polite water-filling levels corresponding to $\tilde{\boldsymbol{\Sigma}}_{1:L}$. 2) By Theorem 13 in [1], $\tilde{\boldsymbol{\Sigma}}_{1:L}$ satisfying the polite water-filling structure also satisfy the KKT conditions of problem (26). Therefore, it is the optimum of problem (26) when the weighted sum rate $\sum_l^L \tilde{\nu}_l \mathcal{I}_l (\boldsymbol{\Sigma}_{1:L}, \boldsymbol{\Phi})$ is a concave function. The sufficiency for **SPMP** can be proved similarly.

We check whether the solutions of Algorithms A and B satisfy the optimality conditions. We use the notation $^-$ for the variables corresponding to the solution of Algorithm A or B. The following is obvious.

*Lemma 2:* After the convergence of the Algorithm A or B, the following conditions are satisfied.

1) In the forward (reverse) links, the MMSE-SIC receive vectors corresponding to $\bar{\mathbf{T}}$ and $\bar{\mathbf{p}}$ ($\bar{\mathbf{R}}$ and $\bar{\mathbf{q}}$) are given by $\bar{\mathbf{R}}$ ($\bar{\mathbf{T}}$). The set of SINRs achieved by $\{\bar{\mathbf{T}}, \bar{\mathbf{R}}, \bar{\mathbf{p}}\}$ in the forward links equals to that achieved by $\{\bar{\mathbf{R}}, \bar{\mathbf{T}}, \bar{\mathbf{q}}\}$ in the reverse links.

2) For Algorithm B, the achieved rates satisfy $\mathcal{I}_l(\bar{\boldsymbol{\Sigma}}_{1:L}, \boldsymbol{\Phi}) = \mathcal{I}_l^0, l = 1, ..., L$.

3) For Algorithm A, $\bar{\boldsymbol{\Sigma}}_{1:L}$ satisfies $\sum_{l=1}^{L} \text{Tr}(\bar{\boldsymbol{\Sigma}}_l) = P_T$.

*Remark 1:* The rates achieved by Algorithm A may not satisfy the condition $\mathcal{I}_l(\bar{\boldsymbol{\Sigma}}_{1:L}, \boldsymbol{\Phi}) = \alpha \mathcal{I}_l^0, \forall l$. To discuss the optimality, we modify the target rates in **FOP/FOPa** to $\mathcal{I}_l^0 = \mathcal{I}_l(\bar{\boldsymbol{\Sigma}}_{1:L}, \boldsymbol{\Phi})$. Then, we can claim that the solution of Algorithm A also satisfies the second optimality condition. The rest is to check whether $\bar{\boldsymbol{\Sigma}}_{1:L}$ possesses the polite water-filling structure. One might conjecture that the first condition on MMSE structure in Lemma 2 implies the polite water-filling structure. Unfortunately, this is not always true according to the following counter example. Consider a single user channel $\mathbf{H}$ with $\text{Rank}(\mathbf{H}) > 1$ and unequal singular values. If the transmit vectors are initialized as the non-zero right singular vectors of $\mathbf{H}$, the algorithm will converge to a solution where the transmit and receive vectors respectively are the non-zero right and left singular vectors of $\mathbf{H}$, and the transmit powers will make the SINRs of all streams the same. Then the solution does not satisfy the single-user water-filling structure. However, for a smaller class of channels, the MMSE structure in Lemma 2 does imply the polite water-filling structure.

*Theorem 9:* If $\text{Rank}(\mathbf{H}_{l,l}) = 1, \forall l$, the solution of Algorithm A (B) $\bar{\boldsymbol{\Sigma}}_{1:L}$ satisfies all the optimality conditions in Theorem 8, and thus achieves a stationary point.

The proof is given in [23]. For general cases, the solution of Algorithms A or B may not possess the polite water-filling structure. In the next sub-section, we design polite water-filling based algorithms which find solutions satisfying all the optimality conditions in Theorem 8.

### D. Polite Water-filling based Algorithms

We only present the detailed algorithms for **SPMP**. The algorithms for **FOP** are similar and will be briefly discussed. We first propose a monotonically convergent iterative algorithm for a sub-class of B-MAC networks named iTree Networks. Then the algorithm is modified for general B-MAC networks.

*1) Algorithm PR for iTree Networks:* iTree networks defined in [1] appears to be a natural extension of MAC and BC. We review its definition below.

*Definition 3:* A B-MAC network with a fixed coupling matrix is called an *Interference Tree (iTree) Network* if after interference cancellation, the links can be indexed such that any link is not interfered by the links with smaller indices.

We give an example in Fig. 2 where DPC and SIC are employed. With encoding/decoding order A, where the signal $\mathbf{x}_2$ is decoded after $\mathbf{x}_1$ and the signal $\mathbf{x}_3$ is encoded after $\mathbf{x}_2$, each link $l \in \{2, 3, 4\}$ is not interfered by the first $l-1$ links. Therefore, the network in Fig. 2 is an iTree network even though it has a loop. However, it is not an iTree network for encoding/decoding order B, where SIC is not employed at $R_1/R_2$, and $\mathbf{x}_2$ is encoded after $\mathbf{x}_3$ at $T_2/T_3$.

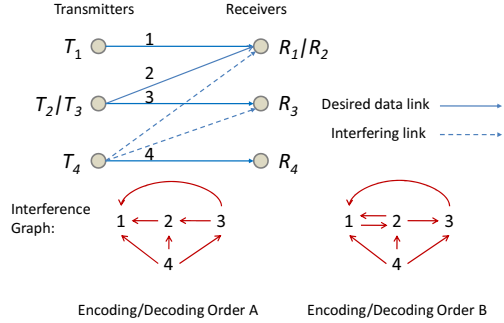

Figure 2.  Illustration of iTree networks

Without loss of generality, we consider iTree networks where the $l^{th}$ link is not interfered by the first $l-1$ links. The following lemma is obvious.

*Lemma 3:* [1] If in an iTree network, the $l^{th}$ link is not interfered by the links with lower indices, in the reverse links, the $l^{th}$ link is not interfered by the links with higher indices.

We show that if any $\mathbf{\Sigma}_i$ does not satisfy the polite water-filling structure, the objective (cost) in **FOP** (**SPMP**) can be strictly increased (decreased) by enforcing this structure at link $i$, which is the key component of Algorithm PR.

We first define some notations. Suppose $\mathbf{\Sigma}_{1:L}$ achieves a rate point $[\mathcal{I}_l]_{l=1,...,L}$ with sum power $P_T \triangleq \sum_{l=1}^{L} \text{Tr}(\mathbf{\Sigma}_l)$ and its covariance transformation $\hat{\mathbf{\Sigma}}_{1:L}$ achieves a rate point $\left[\hat{\mathcal{I}}_l \geq \mathcal{I}_l\right]_{l=1,...,L}$. Fixing $\mathbf{\Sigma}_j$, $j = i+1,...,L$ for the last $L-i$ links, the first $i$ links form a sub-network

$$\left([\mathbf{H}_{l,k}]_{k,l=1,...,i}, \sum_{l=1}^{i} \text{Tr}(\mathbf{\Sigma}_l) = P_T^i, [\mathbf{W}_l]_{l=1,...i}\right), \quad (27)$$

where $\mathbf{W}_l = \mathbf{I} + \sum_{j=i+1}^{L} \mathbf{\Phi}_{l,j} \mathbf{H}_{l,j} \mathbf{\Sigma}_j \mathbf{H}_{l,j}^{\dagger}$, $\forall l$ is the covariance matrix of the equivalent colored noise; $P_T^i = \sum_{l=1}^{i} \text{Tr}(\mathbf{\Sigma}_l)$. The dual sub-network is

$$\left(\left[\mathbf{H}_{k,l}^{\dagger}\right]_{k,l=1,...,i}, \sum_{l=1}^{i} \text{Tr}\left(\hat{\mathbf{\Sigma}}_l \mathbf{W}_l\right) = P_T^i, [\mathbf{I}]_{l=1,...,i}\right). \quad (28)$$

By Lemma 9 in [1], $\hat{\mathbf{\Sigma}}_{1:i} = \left(\hat{\mathbf{\Sigma}}_1,...,\hat{\mathbf{\Sigma}}_i\right)$ is also the covariance transformation of $\mathbf{\Sigma}_{1:i} = (\mathbf{\Sigma}_1,...,\mathbf{\Sigma}_i)$, applied to the sub-network (27).

The performance can be improved as follows.

**Step** 1: Improve the rate of reverse link $i$ by enforcing the polite water-filling structure on $\hat{\mathbf{\Sigma}}_i$. By Lemma 3, the reverse link $i$ causes no interference to the first $i-1$ reverse links and thus its rate can be improved without hurting other reverse links in the sub network by solving the following single-user optimization problem:

$$\max_{\hat{\boldsymbol{\Sigma}}_i \geq 0} \log \left| \mathbf{I} + \mathbf{H}_{i,i}^{\dagger} \hat{\boldsymbol{\Sigma}}_i \mathbf{H}_{i,i} \hat{\boldsymbol{\Omega}}_i^{-1} \right| \quad (29)$$

$$\text{s.t. } \mathrm{Tr}\left(\hat{\boldsymbol{\Sigma}}_i \mathbf{W}_i\right) \leq P_T^i - \sum_{l=1}^{i-1} \mathrm{Tr}\left(\hat{\boldsymbol{\Sigma}}_l \mathbf{W}_l\right),$$

where $\hat{\boldsymbol{\Omega}}_i = \mathbf{I} + \sum_{k=1}^{i-1} \boldsymbol{\Phi}_{k,i} \mathbf{H}_{k,i}^{\dagger} \hat{\boldsymbol{\Sigma}}_k \mathbf{H}_{k,i}$ and $\mathbf{W}_i = \mathbf{I} + \sum_{j=i+1}^{L} \boldsymbol{\Phi}_{i,j} \mathbf{H}_{i,j} \boldsymbol{\Sigma}_j \mathbf{H}_{i,j}^{\dagger} = \boldsymbol{\Omega}_i$. By a simple extension of the results in [30], it can be proved that the optimal solution is uniquely given by the following polite water-filling procedure. Perform the thin SVD $\boldsymbol{\Omega}_i^{-1/2} \mathbf{H}_{i,i} \hat{\boldsymbol{\Omega}}_i^{-1/2} = \mathbf{F}_i \boldsymbol{\Delta}_i \mathbf{G}_i^{\dagger}$. Let $N_i = \mathrm{Rank}\left(\mathbf{H}_{i,i}\right)$ and $\delta_{i,j}$ be the $j^{th}$ diagonal element of $\boldsymbol{\Delta}_i^2$. Obtain $\mathbf{D}_i$ as

$$\begin{aligned} \mathbf{D}_i &= \mathrm{diag}\left(d_{i,j}, ..., d_{i,N_i}\right), \quad (30) \\ d_{i,j} &= \left(\nu_i - \frac{1}{\delta_{i,j}}\right)^+, j = 1, ..., N_i, \end{aligned}$$

where $\nu_i$ is chosen such that $\sum_{j=1}^{N_i} d_{i,j} = P_T^i - \sum_{l=1}^{i-1} \mathrm{Tr}\left(\hat{\boldsymbol{\Sigma}}_l \mathbf{W}_l\right)$. Then the optimal solution is

$$\hat{\boldsymbol{\Sigma}}_i' = \boldsymbol{\Omega}_i^{-1/2} \mathbf{F}_i \mathbf{D}_i \mathbf{F}_i^{\dagger} \boldsymbol{\Omega}_i^{-1/2}. \quad (31)$$

By Theorem 3, if $\boldsymbol{\Sigma}_i$ does not satisfy the polite water-filling, nor does $\hat{\boldsymbol{\Sigma}}_i$, which implies that $\hat{\boldsymbol{\Sigma}}_i'$ achieves a rate $\hat{\mathcal{I}}_i' > \hat{\mathcal{I}}_i$.

**Step** 2: Improve the forward links by the covariance transformation from $\hat{\boldsymbol{\Sigma}}_{1:i}' = \left(\hat{\boldsymbol{\Sigma}}_1, ..., \hat{\boldsymbol{\Sigma}}_{i-1}, \hat{\boldsymbol{\Sigma}}_i'\right)$ to $\boldsymbol{\Sigma}_{1:i}' = \left(\boldsymbol{\Sigma}_1', ..., \boldsymbol{\Sigma}_i'\right)$ for the sub-network[2]. By Theorem 1, the covariance transformation $\boldsymbol{\Sigma}_{1:i}'$ achieves a set of rates satisfying $\mathcal{I}_i' \geq \hat{\mathcal{I}}_i' > \mathcal{I}_i$ and $\mathcal{I}_l' \geq \hat{\mathcal{I}}_l \geq \mathcal{I}_l, l < i$ in the sub-network under the sum power constraint $\sum_{l=1}^{i} \mathrm{Tr}\left(\boldsymbol{\Sigma}_l'\right) = P_T^i$. Since the first $i$ links cause no interference to all other links in the original network, the input covariance matrices $\boldsymbol{\Sigma}_{1:L}' = \left(\boldsymbol{\Sigma}_1', \cdots, \boldsymbol{\Sigma}_i', \boldsymbol{\Sigma}_{i+1}, \cdots, \boldsymbol{\Sigma}_L\right)$ must achieve a rate point satisfying $\mathcal{I}_i' > \mathcal{I}_i$ and $\mathcal{I}_l' \geq \mathcal{I}_l, \forall l \neq i$ with the same sum power $P_T$.

The objective functions of **FOP** and **SPMP** can be strictly improved by the above two steps. Here, we only show how to do it for **SPMP**. Note that the polite water-filling level $\nu_i$ in (30) is chosen to satisfy the forward sum power constraint and improve the rate of reverse link $i$. If the initial solution is feasible, i.e., $\mathcal{I}_i \geq \mathcal{I}_i^0$, we can reduce the polite water-filling level $\nu_i$ to make the rate of reverse link $i$ equal to $\mathcal{I}_i^0$, and thus reduce the forward sum power. This results in an algorithm which monotonically

---

[2]Due to the special interference structure of iTree networks, the calculation of the transmit powers of the covariance transformation can be simplified to be calculated one by one as shown in [1].

Table I

Algorithm W (Solving the Polite Water-filling Level for the Rate Constraints)

1. Initialize the set of indices of the streams of link $i$ as
   $\Gamma = \{1, ..., N_i\}$, where $N_i = \text{Rank}(\mathbf{H}_{i,i})$.
2. Calculate $\nu_i = \left(e^{\mathcal{I}_i^0}/\Pi_{j \in \Gamma} \delta_{i,j}\right)^{1/|\Gamma|}$, which is the solution of
   $\sum_{j \in \Gamma} \log\left(1 + (\nu_i - 1/\delta_{i,j})\delta_{i,j}\right) = \mathcal{I}_i^0$.
   Obtain $d_{i,j} = \nu_i - 1/\delta_{i,j}$ for $j \in \Gamma$.
3. If $d_{i,j} \geq 0, \forall j \in \Gamma$, stop. Otherwise, for all $j \in \Gamma$, if $d_{i,j} < 0$,
   fix it as $d_{i,j} = 0$, delete $j$ from $\Gamma$. Repeat step 2).

Table II

Algorithm PR (Solving **SPMP** for iTree Networks)

Initialize $\mathbf{\Sigma}_{1:L}$ such that $\mathbf{\Sigma}_i \succeq 0, \forall i$.
**While** not converge **do**
 **For** $i = 1 : L$
 1. Calculate $\hat{\mathbf{\Sigma}}_{1:i}$ by the covariance transformation of $\mathbf{\Sigma}_{1:i}$
    applied to the $i^{\text{th}}$ sub-network.
 2. Obtain $\hat{\mathbf{\Sigma}}'_i$ by polite water-filling as in (30) and (31), where
    the polite water-filling level $\nu_i$ is calculated by Algorithm W.
 3. Calculate $\mathbf{\Sigma}'_{1:i}$ by the covariance transformation of
    $\hat{\mathbf{\Sigma}}'_{1:i} = \left(\hat{\mathbf{\Sigma}}_1, ..., \hat{\mathbf{\Sigma}}_{i-1}, \hat{\mathbf{\Sigma}}'_i\right)$ applied to the $i^{\text{th}}$ sub-network.
 4. Update $\mathbf{\Sigma}_{1:L}$ as $\mathbf{\Sigma}_{1:L} = \left(\mathbf{\Sigma}'_1, ..., \mathbf{\Sigma}'_i, \mathbf{\Sigma}_{i+1}, ..., \mathbf{\Sigma}_L\right)$.
 **End**
**End**

decreases the sum power once the solution becomes feasible. A simple algorithm in Table I referred to as *Algorithm W* can be used to calculate the polite water-filling level $\nu_i$ to satisfy the rate constraint $\mathcal{I}_i^0$.

The overall algorithm for iTree networks is summarized in table II and referred to as Algorithm PR, where P stands for Polite and R stands for Rate constraint. The following theorem is obvious.

*Theorem 10:* Once Algorithm PR finds a feasible solution, it will monotonically decrease the sum power until it converges to a stationary point.

*2) Algorithm PR1 for B-MAC Networks:* We obtain Algorithm PR1 in table III for general B-MAC networks by a modification of Algorithm PR so that the polite water-filling structure is imposed iteratively. The algorithm can also be derived from the Lagrange function of the problem, where the Lagrange

Table III

ALGORITHM PR1 (SOLVING **SPMP** FOR B-MAC NETWORKS)

---

Initialize $\hat{\boldsymbol{\Sigma}}_{1:L}$ and $\boldsymbol{\Omega}_i$'s such that $\hat{\boldsymbol{\Sigma}}_i \succeq 0, \forall i$ and $\boldsymbol{\Omega}_i = \mathbf{I}, \forall i$.

1. Update in the forward links

 a. For $\forall i$, obtain $\hat{\boldsymbol{\Omega}}_i$ from $\hat{\boldsymbol{\Sigma}}_{1:L}$ using (4).

 Perform thin SVD $\boldsymbol{\Omega}_i^{-1/2}\mathbf{H}_{i,i}\hat{\boldsymbol{\Omega}}_i^{-1/2} = \mathbf{F}_i\boldsymbol{\Delta}_i\mathbf{G}_i^\dagger$.

 b. Obtain $\mathbf{D}_i$ by the water-filling in (30), where

 the polite water-filling level $\nu_i$ is calculated by Algorithm W.

 c. Update $\boldsymbol{\Sigma}_i$'s as $\boldsymbol{\Sigma}_i = \hat{\boldsymbol{\Omega}}_i^{-1/2}\mathbf{G}_i\mathbf{D}_i\mathbf{G}_i^\dagger\hat{\boldsymbol{\Omega}}_i^{-1/2}, \forall i$.

2. Update in the reverse links similar as that in the forward links

 a. For $\forall i$, obtain $\boldsymbol{\Omega}_i$ from $\boldsymbol{\Sigma}_{1:L}$ using (1).

 b. For $\forall i$, update $\hat{\boldsymbol{\Sigma}}_i$ as in (30) and (31), where

 the polite water-filling level $\nu_i$ is calculated by Algorithm W.

Return to step 1 until converge.

---

multipliers are exactly the water-filling levels of the links. Adjusting the Lagrange multipliers to satisfy the rate constraints is exactly what Algorithm W does. It is clear that if Algorithm PR1 converges, the solution satisfies the optimality conditions in Theorem 8, and thus achieves a stationary point. The convergence is conjectured and the proof is left for future work. The intuition and simulations strongly indicate fast convergence.

*Remark 2:* Algorithm PR1 can be used to solve the **FOP** by replacing constraints $\mathcal{I}_l^0$ with $\alpha\mathcal{I}_l^0$ and searching for $\alpha$ to satisfy the power constraint.

*Remark 3:* Algorithm PR1 can be easily implemented distributedly as shown in Section IV-E. Another advantage is that it has linear complexity per iteration in terms of the total number of links $L$ [23], while in Algorithm B, the calculation of the transmit powers $\mathbf{p}$ and $\mathbf{q}$ needs to solve two $\sum_{l=1}^{L} M_l$-dimensional linear equations, whose complexity order is usually higher.

*E. Distributed Implementation of Algorithm PR1*

In a network, it is desirable to use distributed optimization. The above centralized algorithms serve as the basis for the distributed design. Here, we design a distributed algorithm based on Algorithm PR1 for time division duplex (TDD) networks. To perform the polite water-filling, $T_l$ ($R_l$) only needs to know the equivalent channel $\boldsymbol{\Omega}_l^{-1/2}\mathbf{H}_{l,l}\hat{\boldsymbol{\Omega}}_l^{-1/2}$, which can be obtained by pilot-aided estimation.

We assume block fading channel, where each block consists of a training stage followed by a transmission stage. The training stage is further divided into rounds, where one round consists of a half round of

Table IV

ALGORITHM PRD (DISTRIBUTED VERSION OF ALGORITHM PR1)

---

Initialize $i = 1$ and $\mathbf{\Omega}_l^{(0)} = \mathbf{I}$, $\hat{\mathbf{\Omega}}_l^{(0)} = \mathbf{I}$, $\forall l$.

1. In the $i^{\text{th}}$ forward training round, $T_l$ calculates $\mathbf{\Sigma}_l^{(i)}$ by polite water-filling over $\left(\mathbf{\Omega}_l^{(i-1)}\right)^{-1/2} \mathbf{H}_{l,l} \left(\hat{\mathbf{\Omega}}_l^{(i-1)}\right)^{-1/2}$ and transmits pilot signals[3]. $R_l$ estimates $\mathbf{\Omega}_l^{(i)}$ and $\mathbf{H}_{l,l} \left(\hat{\mathbf{\Omega}}_l^{(i-1)}\right)^{-1/2}$.

2. In the $i^{\text{th}}$ reverse training round, $R_l$ calculates $\hat{\mathbf{\Sigma}}_l^{(i)}$ by polite water-filling over $\left(\hat{\mathbf{\Omega}}_l^{(i-1)}\right)^{-1/2} \mathbf{H}_{l,l}^\dagger \left(\mathbf{\Omega}_l^{(i)}\right)^{-1/2}$ and transmits pilot signals. $T_l$ estimates $\hat{\mathbf{\Omega}}_l^{(i)}$ and $\mathbf{H}_{l,l}^\dagger \left(\mathbf{\Omega}_l^{(i)}\right)^{-1/2}$.

3. Let $i = i + 1$ and enter the next round. Keep updating $\mathbf{\Sigma}_l^{(i)}$ and $\hat{\mathbf{\Sigma}}_l^{(i)}$ until the end of the training stage.

---

pilot aided estimation of $\mathbf{H}_{l,l}\hat{\mathbf{\Omega}}_l^{-1/2}$ and $\mathbf{\Omega}_l$ in the forward link and a half round of pilot aided estimation of $\mathbf{H}_{l,l}^\dagger \mathbf{\Omega}_l^{-1/2}$ and $\hat{\mathbf{\Omega}}_l$ in the reverse link. The $T_l$'s and $R_l$'s run a distributed version of Algorithm PR1 to solve **SPMP** and use the resulted input covariance matrices for the transmission stage.

First, we describe the operation at $T_l$. The operation at $R_l$ is similar.

- In the $(i-1)^{\text{th}}$ reverse training round, $T_l$ estimates the interference-plus-noise covariance matrix $\hat{\mathbf{\Omega}}_l^{(i-1)}$ and the effective channel $\mathbf{H}_{l,l}^\dagger \left(\mathbf{\Omega}_l^{(i-1)}\right)^{-1/2}$.
- In the $i^{\text{th}}$ forward training round, $T_l$ calculates the input covariance matrix $\mathbf{\Sigma}_l^{(i)}$ by polite water-filling over the equivalent channel $\left(\mathbf{\Omega}_l^{(i-1)}\right)^{-1/2} \mathbf{H}_{l,l} \left(\hat{\mathbf{\Omega}}_l^{(i-1)}\right)^{-1/2}$ as in step 1 of Algorithm PR1. If $\text{Tr}\left(\mathbf{\Sigma}_l^{(i)}\right) > P_{T_l}^{\max}$, where $P_{T_l}^{\max}$ is the maximum transmit power of $T_l$, decrease the polite water-filling level $\nu_l$ until $\text{Tr}\left(\mathbf{\Sigma}_l^{(i)}\right) = P_{T_l}^{\max}$. Then $T_l$ transmits pilot signals.

We summarize this distributed algorithm called PRD in Table IV

Since one training round is almost the same as one iteration in Algorithm PR1 except that each node has an additional maximum power constraint, Algorithm PRD achieves nearly the same performance as Algorithm PR1 after convergence. It is observed that very few training rounds, usually 2.5 to 3.5, suffices for Algorithm PRD to achieve most of the gain, a desirable property for practical applications.

*F. Optimization of the Encoding and Decoding Order*

When DPC and SIC are employed, the coupling matrix $\mathbf{\Phi}(\pi)$ is a function of the encoding and decoding order $\pi$. Finding the optimal $\pi$ is generally difficult because the encoding and decoding orders at the BC transmitters and the MAC receivers need to be solved jointly. However, for each *Pseudo BC*/*Pseudo MAC* defined below, the optimal $\pi$ is characterized in Theorem 11.

*Definition 4:* In a B-MAC network, a set of links, whose indices forms a set $\mathcal{L}_{\text{B}}$, associated with a single physical transmitter is said to be a *Pseudo BC* if either all links in $\mathcal{L}_{\text{B}}$ completely interfere with a link $k$ or all links in $\mathcal{L}_{\text{B}}$ do not interfere with a link $k$, $\forall k \in \mathcal{L}_{\text{B}}^C$. A set of links, whose indices forms a set $\mathcal{L}_{\text{M}}$, associated with a single physical receiver is said to be a *Pseudo MAC* if either all links in $\mathcal{L}_{\text{M}}$ are completely interfered by a link $k$ or all links in $\mathcal{L}_{\text{M}}$ are not interfered by a link $k$, $\forall k \in \mathcal{L}_{\text{M}}^C$.

*Example 1:* In Fig. 1, suppose $\mathbf{x}_1$ is encoded after $\mathbf{x}_2$ and $\mathbf{x}_4$ is the last one to be decoded at the second physical receiver. Then link 2 and link 3 form a pseudo MAC because they belong to the same physical receiver and suffer the same interference from $\mathbf{x}_1$, $\mathbf{x}_4$ and $\mathbf{x}_5$. Similarly, link 4 and link 5 form a pseudo BC.

*Remark 4:* The pseudo BC and pseudo MAC were first introduced in [1] where the optimal encoding/decoding order for the weighted sum-rate maximization problem (**WSRMP**) is shown to be consistent with that of an individual BC or MAC. Similar results are obtained for **FOP** and **SPMP** in Theorem 11.

First, we need to modify the **FOP** and **SPMP** to include encoding and decoding order optimization and time sharing. Let $\Xi$ be a set of valid coupling matrices produced by proper encoding and decoding orders. Define a larger achievable region

$$\mathcal{R}(P_T) \;=\; \text{Convex Closure} \bigcup_{\boldsymbol{\Phi} \in \boldsymbol{\Xi}} \mathcal{R}_{\boldsymbol{\Phi}}(P_T).$$

The modified optimization problems are

$$\textbf{OFOP}: \max_{\mathbf{r} \in \mathcal{R}(P_T)} \left( \min_{1 \le l \le L} \frac{r_l}{\mathcal{I}_l^0} \right), \tag{32}$$

and

$$\textbf{OSPMP}: \min_{P_T} \; P_T \tag{33}$$
$$\text{s.t.} \quad [\mathcal{I}_l^0]_{l=1,\ldots,L} \in \mathcal{R}(P_T).$$

The following lemma is a consequence of that all outer boundary points of $\mathcal{R}(P_T)$ are Pareto optimal and can be proved by contradiction.

*Lemma 4:* The optimal solution of **OFOP** or **OSPMP** is the intersection of the ray $\alpha \left[ \mathcal{I}_1^0, \cdots, \mathcal{I}_L^0 \right]$, $\alpha > 0$, and the boundary of $\mathcal{R}(P_T)$, where for **OSPMP**, the sum power $P_T$ is chosen such that the intersection is at $\left[ \mathcal{I}_1^0, \cdots, \mathcal{I}_L^0 \right]$.

The following theorem characterizes the optimal encoding and decoding order for those boundary points of $\mathcal{R}(P_T)$ that can be achieved by DPC and SIC without time sharing.

*Theorem 11:* Necessity: If the input covariance matrices $\tilde{\mathbf{\Sigma}}_{1:L}$ and a valid encoding and decoding order $\tilde{\pi}$ achieves the optimum of **OFOP** and **OSPMP** without time sharing, they must satisfy the following necessary conditions:

1) $\tilde{\mathbf{\Sigma}}_{1:L}$ satisfies the optimality conditions in Theorem 8.
2) If there exists a pseudo BC (pseudo MAC) in the B-MAC network, its optimal encoding (decoding) order satisfies that, the signal of the link with the $n^{\text{th}}$ largest (smallest) polite water-filling level is the $n^{\text{th}}$ one to be encoded (decoded).

Sufficiency: In MAC or BC, if certain $\tilde{\mathbf{\Sigma}}_{1:L}$ and $\tilde{\pi}$ satisfy the above conditions, they must be the optimum of **OFOP** or **OSPMP**.

*Proof:* The first necessary condition follows from Theorem 8. The second necessary condition follows from the following two facts and Lemma 4. 1) Any outer boundary point of $\mathcal{R}(P_T)$ must be the solution of a **WSRMP** with certain weight vector $[w_l]_{l=1,\ldots,L}$. It is proved in [1] that the optimal solution of a **WSRMP** must satisfy the polite water-filling structure and the polite water-filling levels are given by $\nu w_l$'s for some constant $\nu > 0$; 2) By Theorem 9 in [1], the weighted sum-rate optimal encoding and decoding order of each Pseudo BC (Pseudo MAC) is that the signal of the link with the $n^{\text{th}}$ largest (smallest) weight is the $n^{\text{th}}$ one to be encoded (decoded). The sufficiency part is proved as follows. For MAC, certain $\tilde{\mathbf{\Sigma}}_{1:L}$ and $\tilde{\pi}$ satisfy the two conditions in Theorem 11 implies that $\tilde{\mathbf{\Sigma}}_{1:L}$ and $\tilde{\pi}$ maximizes the concave weighted sum-rate $\sum_{l}^{L} \tilde{\nu}_l \mathcal{I}_l \left( \mathbf{\Sigma}_{1:L}, \mathbf{\Phi}(\pi) \right)$ and thus achieves a boundary point of the capacity region of MAC. By Lemma 4, they achieve the global optimality of **OFOP** or **OSPMP**. The sufficiency part for BC follows from the rate duality. ∎

The above proof suggests an algorithm to adjust the encoding/decoding order $\pi$ according to current polite water-filling levels. It is referred to as Algorithm O and summarized in Table V.

*Remark 5:* For the special cases of MAC and BC, if Algorithm O converges, the solution gives the optimal order by Theorem 11.

*Remark 6:* A simple sub-optimal algorithm solving **OSPMP** for the special case of SIMO MAC/MISO BC has been proposed in [16]. The difference between Algorithm O and the sub-optimal algorithm are as follows. 1) The sub-optimal algorithm works for SIMO MAC, avoiding the calculation of beamforming matrices, while Algorithm O works for MIMO cases; 2) In the sub-optimal algorithm, the update of $\pi$ is more complicated, while in Algorithm O, $\pi$ is directly determined by the polite water-filling levels. Same as the sub-optimal algorithm in [16], Algorithm O may cycle through a finite number of orders.

Table V

ALGORITHM O (IMPROVING THE ENCODING AND DECODING ORDER)

---

Initialize the encoding and decoding order $\pi$ such that $\Phi(\pi)$ is valid.
1. Solve the **FOP** or **SPMP** with fixed $\Phi(\pi)$.
2. For each Pseudo BC and Pseudo MAC with the polite water-filling levels obtained in step 1, if $\pi$ satisfies Theorem 11, output $\pi$ and stop. Otherwise, set $\pi$ to satisfy Theorem 11 and return to step 1.

---

In this case, we can choose the best one of them. It is observed that the reason of non-convergence is usually that the corresponding boundary point cannot be achieved without time-sharing.

## V. SIMULATION RESULTS

Simulations are used to demonstrate the performance of the proposed algorithms. Let each transmitter and receiver equipped with $L_T$ and $L_R$ antennas respectively. DPC and SIC are employed for interference cancellation. Block fading channel is assumed and the channel matrices are independently generated by $\mathbf{H}_{l,k} = \sqrt{g_{l,k}} \mathbf{H}_{l,k}^{(W)}, \forall k, l$, where $\mathbf{H}_{l,k}^{(W)}$ has zero-mean i.i.d. Gaussian entries with unit variance; and $g_{l,k}, \forall k, l$ is set as 0dB except for Fig. 4 and Fig. 5. In Fig. 6-8, each simulation is averaged over 100 channel realizations, while in other figures, a single channel realization is considered. We call *pseudo global optimum* the best solution among many solutions obtained by running the algorithm for many times with different initial points and with the encoding/decoding order obtained by Algorithm O. For the plots with iteration numbers, we show rates or power after $x.5$ iterations/rounds, where the last 0.5 iteration/round is the forward link update.

Algorithm A can be used to find the achievable rate region boundary by varying the target rates $\mathcal{I}_l^0$'s. It finds the point where the boundary is intersected by the ray $\alpha \left( \mathcal{I}_1^0, \cdots, \mathcal{I}_L^0 \right)$. In Fig. 3, we plot the boundaries of the rate regions achieved by Algorithm A with different decoding orders for a two-user MAC with $L_T = 2$, $L_R = 4$. It can be observed that the convex hull of the rate regions achieved by Algorithm A is the same as the capacity region, which implies that Algorithm A achieves the optimum for this case, and thus is a low complexity approach to calculate the capacity region for MIMO MAC.

Algorithm PR and PR1 have superior convergence speed. Sum power versus iteration number is shown in Fig. 4 for the iTree network of Fig. 2 and in Fig. 5 for a 3-user interference channel. Each node has four antennas. The target rate of each link is 5 bits/channel use. In the upper sub-plot of Fig. 4, we consider the moderate interference case, where $g_{l,k} = 0$dB, $\forall k, l$. In the lower sub-plot, we consider strong interference case, where $g_{l,3} = 10$dB, $l = 1, 2$ for the interfering links, and $g_{l,k} = 0$dB for other $k, l$'s. It is not surprising that Algorithms PR and PR1 have faster convergence speed because polite water-filling exploits the structure of the problem. In the upper sub-plot of Fig. 5, we set $g_{l,k} = 0$dB, $\forall k, l$. In the lower sub-plot, we consider a strong interference channel with $g_{l,k} = 10$dB, $\forall k \neq l$, and $g_{l,k} = 0$dB, $\forall k = l$. Again, Algorithm PR1 converges faster than Algorithm B.

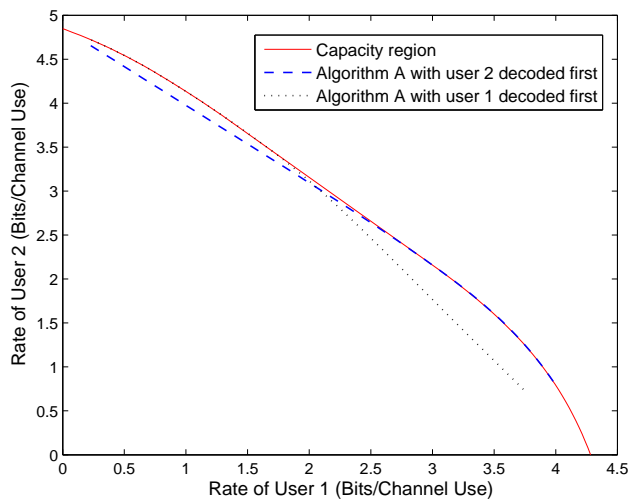

Figure 3. Achieved rate region boundaries of a two-user MAC

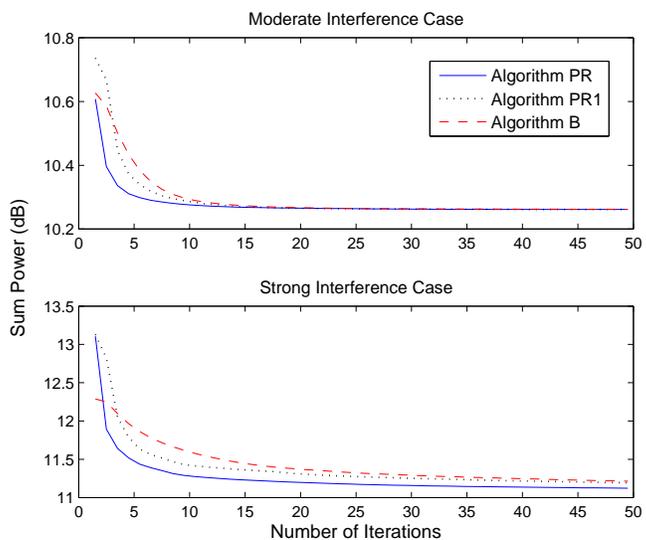

Figure 4. Convergence of the algorithms for the iTree network in Fig. 2

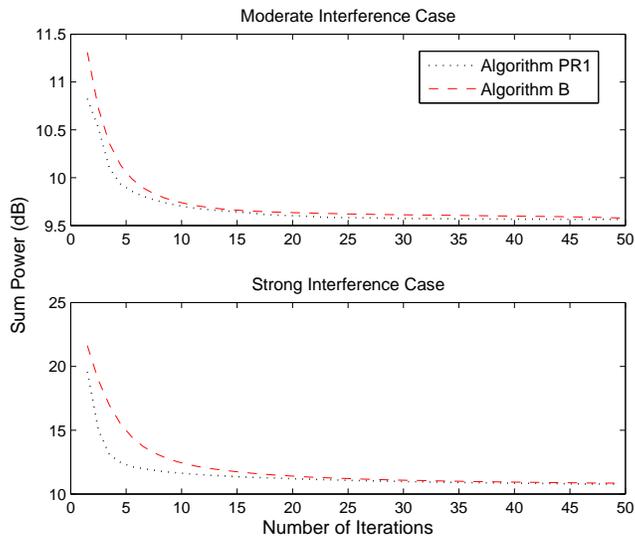

Figure 5. Convergence of the algorithms for a 3-user interference channel

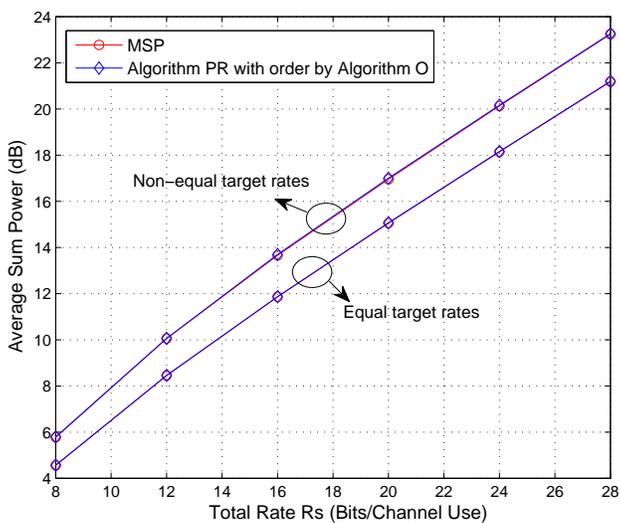

Figure 6. Average sum power vs. the total rate for a 4-user MAC

In Fig. 6, we evaluate the performance of Algorithm PR for a 4-user MAC with $L_T = 2$, $L_R = 4$. The 'MSP' is the optimal solution obtained by the 'Algorithm 2' in [15], which has much higher complexity as discussed in Section I-B. In both equal ($[R_s, R_s, R_s, R_s]/4$) and unequal ($[R_s, 2R_s, 4R_s, 8R_s]/15$) target rate cases, where $R_s$ is the total rate required, Algorithm PR with the decoding order obtained by Algorithm O achieves nearly the same sum power as the MSP but with much lower complexity. Not showing is that Algorithm B and PR1 also achieve the same performance as Algorithm PR. In Fig. 7, we evaluate the performance of Algorithms B and PR1 for the B-MAC network in Fig. 1 with $L_T = L_R = 4$. The target rates are set as $[R_s, R_s, 2R_s, 4R_s, 8R_s]/16$. The encoding/decoding order is partially fixed and is the same as that in Example 1 of Section IV-F. For the pseudo MAC formed by link 2 and link 3 and the pseudo BC formed by link 4 and link 5, the fixed order that $\mathbf{x}_3$ is decoded after $\mathbf{x}_2$ and $\mathbf{x}_5$ is decoded after $\mathbf{x}_4$, and its improved order obtained by Algorithm O are applied. With the improved order obtained by Algorithm O, both algorithms achieve nearly the same performance as the Pseudo global optimum of Algorithm PR1, while the algorithms with the fixed order suffers a performance loss.

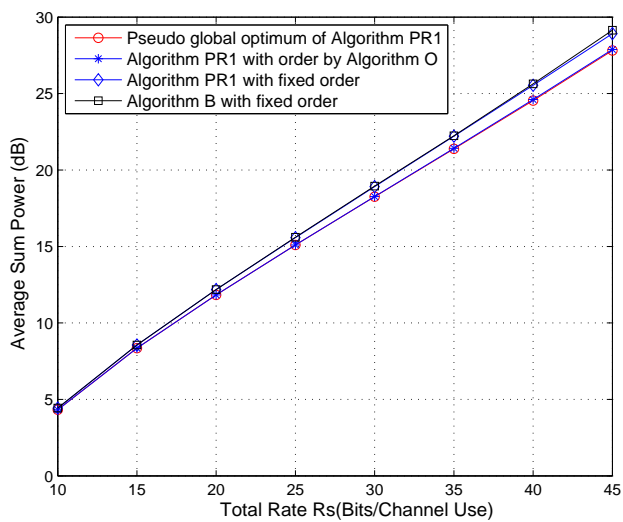

Figure 7. Average sum power vs. the total rate for the B-MAC network in Fig. 1

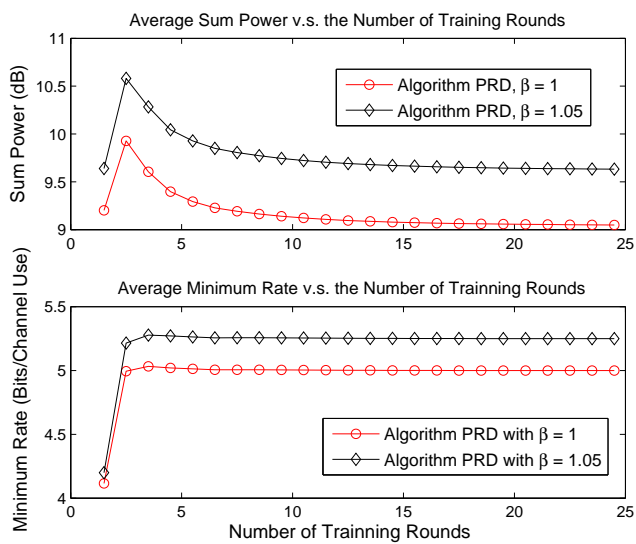

Figure 8. Convergence of the distributed algorithm for a 3-user interference channel

We illustrate the convergence behavior of the distributed algorithm. Fig. 8 plots the total transmit power and the minimum rate of the users achieved by Algorithm PRD versus the number of training rounds for a 3-user interference channel with $L_T = L_R = 4$. Let $\left(\mathcal{I}_1^0, \mathcal{I}_2^0, \mathcal{I}_3^0\right)$, $\mathcal{I}_l^0 = 5$ (bits/channel use), $l = 1, 2, 3$, be the true target rates. In Algorithm PRD, we may set a higher target rates as $\beta \left(\mathcal{I}_1^0, \mathcal{I}_2^0, \mathcal{I}_3^0\right)$ with $\beta = 1$ and $\beta = 1.05$ respectively. It can be observed that after 2.5 rounds the rates are close to the targets and after 3.5 rounds, the powers are also close to the that of infinite rounds. When $\beta = 1$, the achieved minimum rate after 3.5 rounds exceeds the target rate in 88 out of 100 simulations. When $\beta = 1.05$, the target rates are always satisfied after 3.5 rounds, while the total transmit power is about 0.7 dB larger. This suggests a trick that use higher target rates than true targets in order to exactly satisfy the rate constraints in fewer number of iterations at the expense of more power.

## VI. CONCLUSION

The general MIMO one-hop interference networks named B-MAC networks with Gaussian input and any valid coupling matrices are considered. We design low complexity and high performance algorithms for maximizing the minimum of weighted rates under sum power constraints and for minimizing sum power under rate constraints. They can be used in admission control and in guaranteeing the quality of service. Two kinds of algorithms are designed. The first kind takes advantage of existing SINR optimization algorithms by finding simple and optimal mappings between the achievable rate region and the SINR region. The mappings can be used for many other optimization problems. The second kind takes advantage of the polite water-filling structure of the optimal input found recently in [1]. Both centralized and distributed algorithms are designed. The proposed algorithms are either proved or shown by simulations to converge to a stationary point, which may not be optimal for non-convex cases, but is shown by simulations to be good solutions.

## APPENDIX

*A. Proof for Theorem 4*

Without loss of generality, we only prove that for any $\boldsymbol{\Sigma}$ achieving a rate $\mathcal{I}(\boldsymbol{\Sigma}) = \log\left|\mathbf{I} + \mathbf{H}\boldsymbol{\Sigma}\mathbf{H}^{\dagger}\right|$ in a single-user channel $\mathbf{H}$, there exists a decomposition $\boldsymbol{\Sigma} = \dot{\mathbf{T}}\dot{\mathbf{T}}^{\dagger}$ leading to $\{\mathbf{T}, \mathbf{R}, \mathbf{p}\}$ which achieves a set of SINRs $\gamma_m = e^{\mathcal{I}(\boldsymbol{\Sigma})/M} - 1, m = 1, ..., M$.

First, we show that considering unitary precoding matrix $\mathbf{V} \in \mathbb{C}^{M \times M}$ will not loss generality. Note that $\mathcal{I}(\boldsymbol{\Sigma}) = \log\left|\mathbf{I} + \mathbf{H}\dot{\mathbf{T}}\mathbf{V}\mathbf{V}^{\dagger}\dot{\mathbf{T}}^{\dagger}\mathbf{H}^{\dagger}\right| = \log\left|\mathbf{I} + \bar{\mathbf{H}}\mathbf{V}\mathbf{V}^{\dagger}\bar{\mathbf{H}}^{\dagger}\right|$, where $\bar{\mathbf{H}} = \mathbf{H}\dot{\mathbf{T}}$ is the equivalent channel with unitary precoding matrix $\mathbf{V} = [\mathbf{v}_1, ..., \mathbf{v}_M]$. Define

$$\bar{\mathbf{A}}_m = \bar{\mathbf{H}}^{\dagger}\left(\sum_{i=m+1}^{M}\bar{\mathbf{H}}\mathbf{v}_i\mathbf{v}_i^{\dagger}\bar{\mathbf{H}}^{\dagger} + \mathbf{I}\right)^{-1}\bar{\mathbf{H}}.$$

The SINR of the $m^{th}$ stream achieved by the MMSE-SIC receiver is given by [27] $\gamma_m = \mathbf{v}_m^{\dagger}\bar{\mathbf{A}}_m\mathbf{v}_m$. Hence, we only need to find a unitary precoding matrix $\mathbf{V}$ such that $\mathbf{v}_m^{\dagger}\bar{\mathbf{A}}_m\mathbf{v}_m = e^{\mathcal{I}(\boldsymbol{\Sigma})/M} - 1, m = 1, ..., M$. Then the precoding matrix for the original channel is give by $\dot{\mathbf{T}}' = \dot{\mathbf{T}}\mathbf{V}$.

We will use the method of induction. We first find a unit vector $\mathbf{v}_M$ such that $\mathbf{v}_M^{\dagger}\bar{\mathbf{A}}_M\mathbf{v}_M = e^{\mathcal{I}(\boldsymbol{\Sigma})/M} - 1$. Let $\lambda_i^{(M)}$ be the $i^{th}$ largest eigenvalue of $\bar{\mathbf{A}}_M$ and $\mathbf{u}_i^{(M)}$ be the corresponding eigenvector. Since $\mathcal{I}(\boldsymbol{\Sigma}) = \log\left|\mathbf{I} + \bar{\mathbf{H}}\mathbf{V}\mathbf{V}^{\dagger}\bar{\mathbf{H}}^{\dagger}\right| = \log\left|\mathbf{I} + \bar{\mathbf{A}}_M\right| = \log\prod_{i=1}^{M}\left(1 + \lambda_i^{(M)}\right)$, we must have $\lambda_1^{(M)} \geq e^{\mathcal{I}(\boldsymbol{\Sigma})/M} - 1$ and $\lambda_M^{(M)} \leq e^{\mathcal{I}(\boldsymbol{\Sigma})/M} - 1$. Note that $\mathbf{v}_M^{\dagger}\bar{\mathbf{A}}_M\mathbf{v}_M = \sum_{i=1}^{M}\left|\mathbf{v}_M^{\dagger}\mathbf{u}_i^{(M)}\right|^2\lambda_i^{(M)}$. Because $\left\{\mathbf{u}_i^{(m)}, i = 1, ..., M\right\}$ form orthogonal bases, there exists a $\mathbf{v}_M$ such that

$$\left|\mathbf{v}_M^{\dagger}\mathbf{u}_i^{(M)}\right|^2 = 0, \ i = 2, ..., M-1,$$

$$\left|\mathbf{v}_M^{\dagger}\mathbf{u}_1^{(M)}\right|^2\lambda_1^{(M)} + \left|\mathbf{v}_M^{\dagger}\mathbf{u}_1^{(M)}\right|^2\lambda_M^{(M)} = e^{\mathcal{I}(\boldsymbol{\Sigma})/M} - 1.$$

Then it follows $\mathbf{v}_M^{\dagger}\bar{\mathbf{A}}_M\mathbf{v}_M = e^{\mathcal{I}(\boldsymbol{\Sigma})/M} - 1$.

Assume we already found a set of mutual orthogonal unit vectors $\mathbf{v}_l, l = m+1, ..., M$ such that $\mathbf{v}_l^{\dagger}\bar{\mathbf{A}}_l\mathbf{v}_l = e^{\mathcal{I}(\boldsymbol{\Sigma})/M} - 1, l = m+1, ..., M$. The rest is to prove that there exists a $\mathbf{v}_m$ such that $\mathbf{v}_m^{\dagger}\bar{\mathbf{A}}_m\mathbf{v}_m = e^{\mathcal{I}(\boldsymbol{\Sigma})/M} - 1$ and $\mathbf{v}_m$ is orthogonal to $\mathbf{v}_l, l = m+1, ..., M$. Perform SVD $\bar{\mathbf{A}}_m = \mathbf{U}_m\mathbf{D}_m\mathbf{U}_m^{\dagger}$. Let $\lambda_n^{(m)}$ be the $n^{th}$ largest eigenvalue of $\bar{\mathbf{A}}_m$ and $\mathbf{u}_n^{(m)}$ be the corresponding eigenvector. Define $\hat{\mathbf{u}}_n^{(m)} = \mathbf{u}_n^{(m)} - \sum_{j=m+1}^{M}\mathbf{v}_j^{\dagger}\mathbf{u}_n^{(m)}\mathbf{v}_j, n = 1, ..., M$, $\hat{\mathbf{U}}_m = \left[\hat{\mathbf{u}}_1^{(m)}, ..., \hat{\mathbf{u}}_M^{(m)}\right]$ and $\tilde{\mathbf{A}}_m = \hat{\mathbf{U}}_m\mathbf{D}_m\hat{\mathbf{U}}_m^{\dagger}$. Then for

SUBMITTED TO IEEE TRANSACTIONS ON SIGNAL PROCESSING, JUN. 2010 28$i, j = 1, ..., m$, we have

$$\begin{aligned}
\mathbf{v}_i^\dagger \bar{\mathbf{A}}_m \mathbf{v}_j &= \mathbf{v}_i^\dagger \sum_{n=1}^M \mathbf{u}_n^{(m)} \lambda_n^{(m)} \left(\mathbf{u}_n^{(m)}\right)^\dagger \mathbf{v}_j \\
&= \mathbf{v}_i^\dagger \sum_{n=1}^M \hat{\mathbf{u}}_n^{(m)} \lambda_n^{(m)} \left(\hat{\mathbf{u}}_n^{(m)}\right)^\dagger \mathbf{v}_j \quad (34) \\
&= \mathbf{v}_i^\dagger \tilde{\mathbf{A}}_m \mathbf{v}_j,
\end{aligned}$$

where (34) follows from the definition of $\hat{\mathbf{u}}_n^{(m)}$ and the fact that $\mathbf{v}_i^\dagger \mathbf{v}_k = 0$, $i = 1, ..., m$, $k = m+1, ..., M$. Because $\tilde{\mathbf{A}}_m$ is positive semidefinite and

$$\tilde{\mathbf{A}}_m \mathbf{v}_i = \mathbf{0}, i = m+1, ..., M, \quad (35)$$

the rank of $\tilde{\mathbf{A}}_m$ must be less than $m+1$. Let $\tilde{\lambda}_i^{(m)}$ be the $i^{th}$ largest eigenvalue of $\tilde{\mathbf{A}}_m$ and $\tilde{\mathbf{u}}_i^{(m)}$ be the corresponding eigenvector. Define $\mathbf{V}_m = [\mathbf{v}_1, ..., \mathbf{v}_m]$. Note that the interference from the last $M - m$ streams is $\sum_{i=m+1}^M \bar{\mathbf{H}} \mathbf{v}_i$. Then the sum rate of the first $m$ streams is given by

$$\log \left| \mathbf{I} + \bar{\mathbf{H}} \mathbf{V}_m \mathbf{V}_m^\dagger \bar{\mathbf{H}}^\dagger \left( \sum_{i=m+1}^M \bar{\mathbf{H}} \mathbf{v}_i \mathbf{v}_i^\dagger \bar{\mathbf{H}}^\dagger + \mathbf{I} \right)^{-1} \right|$$

$$= \log \left| \mathbf{I} + \mathbf{V}_m^\dagger \bar{\mathbf{A}}_m \mathbf{V}_m \right|$$

$$= \log \left| \mathbf{I} + \mathbf{V}_m^\dagger \tilde{\mathbf{A}}_m \mathbf{V}_m \right| \quad (36)$$

$$= \log \left| \mathbf{I} + \mathbf{V}^\dagger \tilde{\mathbf{A}}_m \mathbf{V} \right| \quad (37)$$

$$= \log \prod_{i=1}^m \left(1 + \tilde{\lambda}_i^{(m)}\right) = \frac{m}{M} \mathcal{I}(\boldsymbol{\Sigma}),$$

where (36) and (37) follows from (34) and (35) respectively. Therefore we must have $\tilde{\lambda}_1^{(m)} \geq e^{\mathcal{I}(\boldsymbol{\Sigma})/M} - 1$ and $\tilde{\lambda}_m^{(m)} \leq e^{\mathcal{I}(\boldsymbol{\Sigma})/M} - 1$. Note that

$$\begin{aligned}
\mathbf{v}_m^\dagger \bar{\mathbf{A}}_m \mathbf{v}_m &= \mathbf{v}_m^\dagger \tilde{\mathbf{A}}_m \mathbf{v}_m \\
&= \sum_{n=1}^M \left| \mathbf{v}_m^\dagger \tilde{\mathbf{u}}_n^{(m)} \right|^2 \tilde{\lambda}_n^{(m)}.
\end{aligned}$$

Because $\left\{ \tilde{\mathbf{u}}_i^{(m)}, i = 1, ..., m \right\}$ form orthogonal bases of the $m$-dimensional subspace orthogonal to $\mathbf{v}_l$, $l = m+1, ..., M$, there exits a unit vector $\mathbf{v}_m$ in this subspace such that

$$\left| \mathbf{v}_m^\dagger \tilde{\mathbf{u}}_i^{(m)} \right|^2 = 0, i = 2, ..., m-1, m+1, ..., M,$$

$$\left| \mathbf{v}_m^\dagger \tilde{\mathbf{u}}_1^{(m)} \right|^2 \tilde{\lambda}_1^{(m)} + \left| \mathbf{v}_m^\dagger \tilde{\mathbf{u}}_m^{(m)} \right|^2 \tilde{\lambda}_m^{(m)} = e^{\mathcal{I}(\boldsymbol{\Sigma})/M} - 1.$$

Then we have $\mathbf{v}_m^\dagger \bar{\mathbf{A}}_m \mathbf{v}_m = e^{\mathcal{I}(\boldsymbol{\Sigma})/M} - 1$. This completes the proof.

June 29, 2010 DRAFT

*B. Proof for Theorem 5*

Note that $\boldsymbol{\Sigma} = \sum_{m=1}^{M} p_m \mathbf{t}_m \mathbf{t}_m^{\dagger}$ implies that $M \geq \text{Rank}(\boldsymbol{\Sigma})$. Define an $M \times M$ DFT matrix $\mathbf{F}$ where the element at the $k^{th}$ row and $l^{th}$ column is $\mathbf{F}_{k,l} = e^{-\frac{2\pi k l}{M} j}/\sqrt{M}$. If $M$ is chosen to be greater than or equal to $L_T$, let $\mathbf{F}_0 \in \mathbb{C}^{L_T \times M}$ be the matrix comprised of the first $L_T$ rows of $\mathbf{F}$. Otherwise, let $\mathbf{F}_0 \in \mathbb{C}^{L_T \times M}$ be the matrix such that the upper sub matrix are $\mathbf{F}$, and other elements are zero. Perform SVD $\boldsymbol{\Sigma} = \mathbf{U}\mathbf{D}\mathbf{U}^{\dagger}$, where the diagonal elements of $\mathbf{D}$ are positive and in descending order. Let $\dot{\mathbf{T}} = \mathbf{U}\mathbf{D}^{1/2}\mathbf{F}_0$. It can be verified that $\dot{\mathbf{T}}\dot{\mathbf{T}}^{\dagger} = \boldsymbol{\Sigma}$. The norms of the columns of $\dot{\mathbf{T}}$ are the diagonal elements of $\dot{\mathbf{T}}^{\dagger}\dot{\mathbf{T}} = \mathbf{F}_0^{\dagger}\mathbf{D}\mathbf{F}_0$ and they are equal to $\frac{\sum_{i=1}^{L_T} \mathbf{D}_{i,i}}{M}$. Then the corresponding transmit powers satisfy $p_m = \text{Tr}(\boldsymbol{\Sigma})/M, \forall m$.

## REFERENCES

[1] A. Liu, Y. Liu, H. Xiang, and W. Luo, "Duality, polite water-filling, and optimization for MIMO B-MAC interference networks and iTree networks," *submitted to IEEE Trans. Info. Theory*, Apr. 2010. [Online]. Available: http://arxiv4.library.cornell.edu/abs/1004.2484

[2] M. Maddah-Ali, A. Motahari, and A. Khandani, "Communication over MIMO X channels: Interference alignment, decomposition, and performance analysis," *IEEE Transactions on Information Theory*, vol. 54, no. 8, pp. 3457–3470, Aug. 2008.

[3] S. A. Jafar and S. Shamai, "Degrees of freedom region for the MIMO X channel," *IEEE Transactions on Information Theory*, vol. 54, No. 1, pp. 151–170, Jan. 2008.

[4] V. Cadambe and S. Jafar, "Interference alignment and the degrees of freedom of wireless X networks," *IEEE Transactions on Information Theory*, vol. 55, no. 9, pp. 3893–3908, sept. 2009.

[5] M. Costa, "Writing on dirty paper (corresp.)," *IEEE Trans. Info. Theory*, vol. 29, no. 3, pp. 439–441, 1983.

[6] E. Visotsky and U. Madhow, "Optimum beamforming using transmit antenna arrays," *in Proc. IEEE VTC, Houston, TX*, vol. 1, pp. 851–856, May 1999.

[7] J.-H. Chang, L. Tassiulas, and F. Rashid-Farrokhi, "Joint transmitter receiver diversity for efficient space division multiaccess," *IEEE Transactions on Wireless Communications*, vol. 1, no. 1, pp. 16–27, Jan 2002.

[8] M. Schubert and H. Boche, "Solution of the multiuser downlink beamforming problem with individual SINR constraints," *IEEE Transactions on Vehicular Technology*, vol. 53, no. 1, pp. 18–28, Jan. 2004.

[9] ——, "Iterative multiuser uplink and downlink beamforming under SINR constraints," *IEEE Transactions on Signal Processing*, vol. 53, no. 7, pp. 2324–2334, july 2005.

[10] B. Song, R. Cruz, and B. Rao, "Network duality for multiuser MIMO beamforming networks and applications," *IEEE Trans. Commun.*, vol. 55, no. 3, pp. 618–630, March 2007.

[11] F. Rashid-Farrokhi, K. Liu, and L. Tassiulas, "Transmit beamforming and power control for cellular wireless systems," *IEEE J. Select. Areas Commun.*, vol. Vol. 16, no. 8, pp. 1437–1450, Oct. 1998.

[12] E. Visotski and U. Madhow, "Optimum beamforming using transmit antenna arrays," *in Proc. IEEE VTC, Houston, TX*, vol. 1, pp. 851–856, May 1999.

[13] H. Boche and M. Schubert, "A general duality theory for uplink and downlink beamforming," *in Proc. IEEE Veh. Tech. Conf. (VTC) Fall, Vancouver, Canada*, vol. 1, Sept. 2002.

[14] P. Viswanath and D. Tse, "Sum capacity of the vector Gaussian broadcast channel and uplink-downlink duality," *IEEE Trans. Info. Theory*, vol. 49, no. 8, pp. 1912–1921, 2003.

[15] J. Lee and N. Jindal, "Symmetric capacity of MIMO downlink channels," *Proc. IEEE Int. Symp. Inform. Theory (ISIT), Washington,*, pp. 1031–1035, July 2006.

[16] C.-H. F. Fung, W. Yu, and T. J. Lim, "Precoding for the multiantenna downlink: Multiuser SNR gap and optimal user ordering," *IEEE Transactions on Communications*, vol. 55, no. 1, pp. 188 –197, jan. 2007.

[17] W. Yu, W. Rhee, S. Boyd, and J. Cioffi, "Iterative water-filling for Gaussian vector multiple-access channels," *IEEE Trans. Info. Theory*, vol. 50, no. 1, pp. 145–152, 2004.

[18] N. Jindal, W. Rhee, S. Vishwanath, S. Jafar, and A. Goldsmith, "Sum power iterative water-filling for multi-antenna gaussian broadcast channels," *IEEE Trans. Inform. Theory*, vol. 51, no. 4, pp. 1570–1580, April 2005.

[19] W. Yu, "Sum-capacity computation for the gaussian vector broadcast channel via dual decomposition," *IEEE Trans. Inform. Theory*, vol. 52, no. 2, pp. 754–759, Feb. 2006.

[20] W. Yu, G. Ginis, and J. Cioffi, "Distributed multiuser power control for digital subscriber lines," *IEEE J. Select. Areas Commun.*, vol. 20, no. 5, pp. 1105–1115, 2002.

[21] O. Popescu and C. Rose, "Water filling may not good neighbors make," in *Proceedings of GLOBECOM 2003*, vol. 3, 2003, pp. 1766–1770.

[22] L. Lai and H. El Gamal, "The water-filling game in fading multiple-access channels," *IEEE Transactions on Information Theory*, vol. 54, no. 5, pp. 2110–2122, 2008.

[23] A. Liu, Y. Liu, H. Xiang, and W. Luo, "Technical report: MIMO B-MAC interference network optimization under rate constraints by polite water-filling and duality," *Peking University and University of Colorado at Boulder Joint Technical Report*, Jun 2010. [Online]. Available: http://arxiv4.library.cornell.edu/abs/1006.5445

[24] W. Yu, "Uplink-downlink duality via minimax duality," *IEEE Trans. Info. Theory*, vol. 52, no. 2, pp. 361–374, 2006.

[25] L. Zhang, R. Zhang, Y. Liang, Y. Xin, and H. V. Poor, "On gaussian MIMO BC-MAC duality with multiple transmit covariance constraints," *submitted to IEEE Trans. on Information Theory*, Sept. 2008.

[26] A. Liu, Y. Liu, H. Xiang, and W. Luo, "Polite water-filling for weighted sum-rate maximization in B-MAC networks under multiple linear constraints," pp. 1–10, submitted, Jun. 2010.

[27] M. Varanasi and T. Guess, "Optimum decision feedback multiuser equalization with successive decoding achieves the total capacity of the gaussian multiple-access channel," in *Proc. Thirty-First Asilomar Conference on Signals, Systems and Computers*, vol. 2, 1997, pp. 1405–1409.

[28] A. Liu, Y. Liu, H. Xiang, and W. Luo, "On the duality of the MIMO interference channel and its application to resource allocation," *in Proc. IEEE GLOBECOM '09.*, Dec. 2009.

[29] A. Liu, A. Sabharwal, Y. Liu, H. Xiang, and W. Luo, "Distributed MIMO network optimization based on local message passing and duality," *in Proc. 47th Annu. Allerton Conf. Commun., Contr. Comput., Monticello, Illinois, USA*, 2009.

[30] E. Telatar, "Capacity of multi-antenna gaussian channels," *Europ. Trans. Telecommu.*, vol. 10, pp. 585–595, Nov./Dec. 1999.